\documentclass[a4paper]{article}
\usepackage{graphicx,cite}
\usepackage{dcolumn}
\usepackage{bm}
\usepackage{literat}
\usepackage[left=20mm,top=20mm]{geometry}
\usepackage{amssymb,amsmath}

\newcommand{\D}{{\rm d}}

\title{Properties of nano-pattern formation
in reaction-diffusion systems with hyperbolic transport and multiplicative
noise}
\author{Dmitrii O.~Kharchenko \footnote{dikh@ipfcentr.sumy.ua}, Vasyl O.~Kharchenko,
Sergei V.~Kokhan\\ \textit{Institute of Applied Physics, National Academy of
Sciences of Ukraine},\\ \textit{58 Petropavlovskaya St., 40030 Sumy, Ukraine}}

\begin{document}
\maketitle

\begin{abstract}
We study nano-pattern formation in a stochastic model for adsorption-desorption
processes with interacting adsorbate and hyperbolic transport caused by memory
effects. It is shown that at early stages the system manifests pattern
selection processes. Stationary stable patterns of nano-size are analyzed. It
was found that multiplicative noise satisfying fluctuation-dissipation relation
can induce re-entrant pattern formation related to non-equilibrium transitions.
According to obtained Fokker-Planck equation kinetics of island sizes in a
quasi-stationary limit is discussed. Analytical results are compared with
computer simulations.
\end{abstract}



\section{Introduction}

From theoretical and experimental observations it is known that
reaction-diffusion systems play an important role in the study of generic
spatiotemporal behavior of non-equilibrium systems. Usually such models admit
main contributions related to both local dynamics (chemical reactions type of
birth-and-death processes) and mass transport. Novel experimental methods, such
as field ion microscopy, scanning tunneling microscopy can be used to monitor
chemical reactions on the metal surfaces with atomic resolution.

In adsorption-desorption processes when material can be deposited from the
gaseous phase such experimental methods allow one to investigate formation of
clusters or islands of adsorbed molecules/atoms \cite{ZTWE96}. Such islands can
have linear size of nanometer range \cite{GLRBE94}. In
Refs.\cite{KNSZGC91,PWCL97,BGBK98,CF99,Nature99} it was experimentally shown
that nano-patterns on solid surface and nano-islands in adsorbed mono-atomic
layers can be organized. It was found that lateral interactions between
adsorbate play an important role in processes of pattern formation on metal
surfaces. The adsorbate presence can modify the local crystallographic
structures of the substrate's surface layer producing long range interactions
between adsorbed atoms and their clusters (see for example,
Refs.\cite{ZKRGL94,M91,V92}). It was observed experimentally that
nanometer-sized vacancy islands can be organized in a perfect triangular
lattice when a single monolayer of Ag was exposed on Ru(0001) surface at room
temperature \cite{Nature99}. The origin of such self-organization of vacancy
islands is an elastic interaction between edges of neighboring islands
\cite{Nature99}. Such systems can be widely used in electronic devices in
industrial applications. Despite theoretical studies of nano-patterns were
widely reported (see Refs.\cite{ME94,BHKM97,MW2005,CTT06,CTT07,M2010} and
citations therein) no detailed descriptions of elastic effects leading to
formation vacancy/adsorbate islands lattice were presented.

It is known that such kind of processes can be studied by multiscale modeling
including: \emph{ab-initio} calculations allowing one to set parameters of
primitive cell at zero temperature and some energetic parameters; molecular
dynamics simulations monitoring microscopic processes of interactions of atoms;
Monte-Carlo approach studying physics on bigger time scales. Unfortunately the
corresponding computational time remains excessive and presented methods can be
used to describe the system only on micron scales. To study pattern-forming
transitions one can use mesoscopic approach to modelize self-organization
processes on large sizes. It can be generalized by taking into account
corresponding fluctuating sources representing an influence of a microscopic
level or fluctuations of environment \cite{ME94,BHKM97}.

Properties of pattern formation in systems of adsorption-desorption type with
dissipative dynamics were studied previously \cite{ME94,BHKM97}. Pattern
formation in stochastic systems with internal multiplicative noise obeying
fluctuation-dissipation relation was discussed in
Refs.\cite{MW2005,M2010,PhysD2009}. It was shown that internal multiplicative
noise can sustain stationary patterns of nano-size range; the corresponding
patterning processes are well related to non-equilibrium transitions induced by
internal multiplicative noise (entropy driven phase transitions
\cite{GO2001,CWLe,CIGOC2003,BL2004,EPJ2008}).

Recently modifications of this mesoscopic approach were proposed, where in the
framework of continuous model one can describe processes at microscopic and
diffusion scales. This approach is known in literature as modified phase field
crystals method \cite{EKHG02,EG04,EPBSG07,TG2008}. It can be used to study
crystal growth, solidification processes, motion of defects, phase transitions
in crystalline systems (having special kind of order). Its main advantage is in
introduction of a crystalline order with elastic interactions leading to
formation of crystal lattice and of inertia effects related to a finite
propagation speed \cite{JP89,H1999,RGYAEA10}. Inertial contribution is
essential at early stages in the system dynamics. On the diffusion time scale
one can observe patterning with a formation of crystalline lattice regarding to
the Hook's law. Formally, the elastic effects are modelized by expansion of a
pair interatomic potential in the Fourier space up to the fourth order in
wave-numbers \cite{EKHG02,EG04,Grant2006}. As a result the spatial interactions
are described by the Swift-Hohenberg operator $(q_0^2+\nabla^2)^2$, $\nabla
\equiv
\partial/\partial\mathbf{r}$, $q_0$ is a constant related to  period of
spatial patterns. Influence of both internal and external stochastic sources
onto pattern-forming transitions in crystalline systems were discussed in
Refs.\cite{PhysA2010,CEJP}. It was shown that in such kind of systems due to
finite propagation speed pattern selection processes are possible; the external
noise can principally change topology of patterns realized in a pure
deterministic case.

For class of systems with adsorption-desorption and diffusion processes the
typical deterministic equation, which will interest us is of the form
\begin{equation}\label{eqX}
\partial_t x= f(x)-\nabla\cdot \mathbf{J},
\end{equation}
where $x=x(\mathbf{r},t)$ is the local coverage at surface defined as the
quotient between the number of adsorbed molecules/atoms in a cell of the
surface and the fixed number of available sites in each cell, $x\le 1$. The
term $f(x)$ stands for local dynamics and describes birth-and-death or
adsorption-desorption processes; the flux $\mathbf{J}$ represents the mass
transport.

Most of the theoretical studies deal with the standard Fick law
$\mathbf{J}=-D\nabla x$, $D$ is the diffusion constant. It is known that at
$f=0$ the ordinary diffusion equation $\partial_tx=\nabla\cdot D\nabla x$ has
the unrealistic feature of infinitely fast (infinite) propagation. To avoid
such unphysical situation the diffusion flux can be generalized by taking into
account memory (non-equilibrium) effects
\begin{equation}\label{eqJ}
 \mathbf{J}=-\int\limits_0^tM(t,t')D\nabla x(\mathbf{r},t'){\rm d}t'
\end{equation}
described by the memory kernel $M(t,t')=\tau_J^{-1}\exp(-|t-t'|/\tau_J)$. In
such a case in the limit $f(x)=0$ one gets the finite propagation speed
$\sqrt{D/\tau_J}$. At $\tau_J\to 0$ the assymptotic $M(t,t')=\delta (t-t')$
leads to the Fick law $\mathbf{J}=-D\nabla x$ with infinite propagation. As far
as real systems (molecules, atoms) have finite propagation speed one should use
Eq.(\ref{eqJ}) or the equivalent Maxwell-Cattaneo equation: $\tau_J\partial
_t\mathbf{J}=-\mathbf{J}-D\nabla x$ \cite{H1999}. Equations
(\ref{eqX},\ref{eqJ}) can be combined into one equation of the form
\begin{equation}\label{hypX}
\tau_J\partial^2_{tt} x+(1-\tau_J f'(x))\partial_{t}x= f(x)+\nabla\cdot
D(x)\nabla x,
\end{equation}
where prime denotes derivative with respect to the argument. As was pointed out
in Ref.\cite{H1999} this equation has some restrictions: (i) it typically does
not preserve positivity of the solution $x(\mathbf{r},t)$; (ii) the damping
coefficient must be positive, i.e., $f'(x)<\tau_J^{-1}$.

In this study we are aimed to describe pattern formation processes in system
given by Eq.(\ref{hypX}) generalized by introducing multiplicative noise
satisfying fluctuation-dissipation relation. We use suppositions of the phase
field theory to construct effective model describing formation of nano-patterns
with crystalline order and pattern selection processes. It will be shown that
formation of stationary patterns in stochastic systems is well related to
noise-induced effects, namely, non-equilibrium transitions. We shall argue that
noise correlated in time can induce re-entrant ordering processes, where the
stable or unstable patterns are realized in a window of control parameter
values. Studying behavior of islands size as clusters of dense phase and
vacancy islands of diluted phase we obtain its time asymptotics. Analytical
results are verified by computer simulations.

The paper is organized as follows. In Section 2 we describe the model of our
system. Section 3 is devoted to study stability of patterns at early stages. In
section 4 we derive the effective Fokker-Planck equation to study noise induced
effects in Section 5 and formation of stationary patterns in Section 6. In
Section 7 we discuss evolution of islands size. We prove our analytical results
by computer simulation in Section 8. Finally, conclusions are presented in last
section.

\section{The model}

We consider a model where only one class of particles is possible. Following
Refs.\cite{MW2005,M2010,ME94,BHKM97,CWM2002,HME98_1,HME98_2} one assumes that
the particles can be adsrobed, desorbed, can diffuse and interact among
themselves. Therefore, we introduce the scalar field describing dynamics of the
local coverage at surface $x(\mathbf{r},t)\in[0,1]$. The reaction term
incorporates adsorption and desorption terms as follows $f(x)=k_a
p(1-x)-k_dx\exp(U(\mathbf{r})/T)$. Here $k_a$ and $k_d$ are adsorption and
desorption rates, respectively; $p$ is the partial pressure of the gaseous
phase; $U(\mathbf{r})$ is the interaction potential.

The total flux $\mathbf{J}$ is a sum of both ordinary diffusion flux
$(-D_0\nabla x)$ and flow of adsorbate $(-(D_0/T) x(1-x)\nabla U)$. Here the
multiplier $x(1-x)$ denotes that the flux is only possible to the $(1-x)$ free
sites. Using the phase field crystals methodology \cite{EKHG02,EG04,Grant2006}
for the potential $U$ one assumes
\begin{equation}\label{expansionU}
 U(\mathbf{r})/T=-\int{\rm d}\mathbf{r}' u(\mathbf{r}-\mathbf{r}')x(\mathbf{r}')\simeq
 -\varepsilon x- \mu\nabla^2 x-\varkappa\nabla^4 x,
\end{equation}
where $\varepsilon\equiv T^{-1} \int{\rm d} r u(r)$, $\mu\equiv (1/2T)\int{\rm
d} r r^2u(r)$ and $\varkappa\equiv (1/4T)\int{\rm d} r r^4u(r)$. To estimate
$\varepsilon$, $\mu$ and $\varkappa$ one can suppose that there is maximal
intensity $u(0)=u_{max}$ and characteristic radius $r_0$ \cite{BHKM97},
therefore, one has $\varepsilon\simeq u_{max}r_0^2$, $\mu\simeq \varepsilon
r_0^2$, $\varkappa\simeq\mu r_0^2$\footnote{Here we assume that $r_0\ll
L_{dif}$, where the diffusion length is $L_{dif}=\sqrt{D_0/k_d}$. For $r_0$ one
has estimation $r_0\sim 1nm$, whereas $L_{dif}\sim 1 \mu m$
\cite{HME98_1,HME98_2}.}. The obtained expression can be rewritten in a more
simplest form $U(\mathbf{r})/T\simeq-(1/4\varepsilon)\mathcal{L}_{SH}x$, where
$\mathcal{L}_{SH}\equiv (q_0^2+\nabla^2)^2$ is the Swift-Hohenberg spatial
coupling operator where $q_0^2=\varepsilon/\varkappa$; here we take $\mu=1$
that results to relations: $\varkappa=1/4\varepsilon$, $q_0^2=4\varepsilon^2$.

Next, after substitution of all above terms we get a system of two equations
kind of (\ref{eqX}, \ref{eqJ}) with
\begin{equation}
f(x)=\alpha (1-x)-x\exp(-\varepsilon x),\quad \alpha\equiv k_ap/k_d,
\end{equation}
where the total diffusion flux $\mathbf{J}$ satisfies the following relaxation
equation
\begin{equation}\label{Jtot}
\tau_J\partial_t \mathbf{J}=-\mathbf{J}-D_0[\nabla x-G(x)\nabla
\mathcal{L}_{SH}x];\quad G(x)=\frac{1}{4\varepsilon}x(1-x).
\end{equation}
The equivalent equation for the coverage takes the form
\begin{equation}
\tau_J\partial^2_{tt}x+\gamma(x)\partial_t x=f(x)+D_0\nabla\cdot[\nabla
x-G(x)\nabla \mathcal{L}_{SH}x],
\end{equation}
where $\gamma(x)\equiv 1-\tau_Jf'(x)$. In our stochastic analysis we are based
on this equation generalized by introduction of a multiplicative noise in
\emph{ad hoc} form using the fluctuation dissipation relation. Therefore, the
corresponding Langevin dynamics is described by the equation
\begin{equation}\label{eq8}
\tau_J\partial^2_{tt}x+\gamma(x)\partial_t
x=\varphi(x;\nabla)+g(x)\zeta(\mathbf{r},t).
\end{equation}
Here $\varphi(x;\nabla)\equiv f(x)+D_0\nabla\cdot[\nabla x-G(x)\nabla
\mathcal{L}_{SH}x]$; $\gamma(x)\equiv g^2(x)$ in order to guarantee that the
fluctuation-dissipation relation is satisfied \cite{MPRV}; $\zeta$ represents
Langevin force. From the mathematical viewpoint a generalized Langevin equation
defined with a memory kernel has a noise fulfilling the fluctuation-dissipation
relation with the noise time-correlator as the memory kernel has. Therefore, in
our case the memory kernel is reduced to the delta-function and hence, the
process $\zeta$ is a white noise in time (with a zero correlation time,
$\tau_c=0$).
 It is known that mathematically defined
white noise (with delta-correlation function) has no physical meaning
\cite{Gardiner}, for example: every next value for $\zeta$ does not depend on
it previous value; a spread in values of the noise $\zeta$ and a total power of
the noise are infinite.
 Therefore,
one needs to consider physical fluctuation source having small but finite
correlation time $\tau_c\ll 1$ (see, for example \cite{Hanggi}). Hence, in our
study we consider a generalized model of the Gaussian noise with zero mean and
a correlator
\begin{equation}\label{LE}
\langle\zeta(\mathbf{r},t)\zeta(\mathbf{r}',t')\rangle
=\frac{2\sigma^2}{\tau_c}\exp\left(-\frac{|t-t'|}{\tau_c}\right)\delta(\mathbf{r}-\mathbf{r'}),
\end{equation}
where $\sigma^2$ is the noise intensity. Hence, a set of main functions
defining the system behavior is reduced to the reaction term $f(x)$ and the
field-dependent diffusion coefficient $G(x)$.

\section{Linear stability analysis}

Let us study fluctuation effects at early stages. The stability analysis for
the system can be done considering a behavior of the averaged quantity $\langle
x\rangle$ in the vicinity of the state $x_0$ related to unique minimum position
of a bare potential $V(x)=-\int_0^x{\rm d}x'f(x')$. Averaging Eq.(\ref{eq8})
one gets
\begin{equation}
\tau_J\partial^2_{tt} \langle x\rangle+\left<\gamma(x)\partial_{t}x\right>=
\left<f(x)+D_0\nabla\cdot[\nabla x-G(x)\nabla \mathcal{L}_{SH}x]
\right>+\left<g(x)\zeta(\mathbf{r},t)\right>.
\end{equation}
Making use of the Novikov theorem \cite{Novikov}, for the noise correlator we
have
$\left<g(x)\zeta(\mathbf{r},t)\right>=(\sigma^2/2)\left<\gamma'(x)\right>$.
Therefore, in the vicinity of the state $x_0$ for the Fourier component
$\langle \delta x_\mathbf{k}(\omega)\rangle$ of the quantity $\langle\delta
x\rangle=\langle x(\mathbf{r},t)\rangle -x_0$ the dispersion relation takes the
form
\begin{equation}\label{sag}
- \tau_J\omega^2-{\rm i}\gamma(x_0)\omega =
f'(x_0)-D_0k^2(1-G(x_0)[q_0^2-k^2]^2)-\frac{\tau_J\sigma^2}{2}f'''(x_0).
\end{equation}
It leads to the following dependencies for dispersion curves:
\begin{equation}\label{dr}
\omega(k)_\mp=-\frac{{\rm i}\gamma(x_0)}{2\tau_J}\mp
 \left[
	 \frac{D_0k^2(1-G(x_0)[q_0^2-k^2]^2)-f'(x_0)}{\tau_J}+\frac{\sigma^2}{2}f'''(x_0)-\frac{\gamma^2(x_0)}{4\tau^2_J}\right]^{1/2}.
\end{equation}
One can see that $\omega(k)$ can have real and imaginary parts, i.e.
$\omega(k)=\Re \omega(k)\pm {\rm i}\Im\omega(k)$. The component $\Re \omega(k)$
is responsible for oscillatory solutions, whereas $\Im\omega(k)$ describes a
stability of the solution $\langle \delta x_\mathbf{k}(\omega)\rangle$.
Analysis of both $\Re \omega(k)$ and $\Im\omega(k)$ allows us to set a
threshold for a wave-number where oscillatory solutions are possible, from one
hand, and, from another one, to define a wave-number for the first unstable
solution. From the obtained dispersion relations it follows that at $k=k_0$
satisfying equation
\begin{equation}
k_0^2(1-G(x_0)[q_0^2-k_0^2]^2)=\frac{1}{D_0}\left[f'(x_0)-\frac{\sigma^2\tau_J}{2}f'''(x_0)+\frac{\gamma^2(x_0)}{4\tau_J}\right]
\end{equation}
two branches of the dispersion relation degenerate. The unstable mode appears
at $k=k_c$ obtained from the equation
\begin{equation}\label{kc}
k_c^2(1-G(x_0)[q_0^2-k_c^2]^2)=\frac{1}{D_0}\left[f'(x_0)-\frac{\sigma^2\tau_J}{2}f'''(x_0)\right].
\end{equation}

From Eq.(\ref{sag}) it follows that a sign of the effective control parameter
$\varepsilon_{ef}\equiv f'(x_0)-(\tau_J\sigma^2/2) f'''(x_0)$ defines stability
of the system. Moreover, as far as the sign of $\varepsilon_{ef}$ depends on
the reaction term derivatives in the vicinity of the state $x_0$, one can
conclude that noise-induced effects in the linear regime are possible at
$\tau_J\ne0$ and $f'''(x_0)\ne 0$.

\begin{figure}
\centering
 a)\includegraphics[width=80mm]{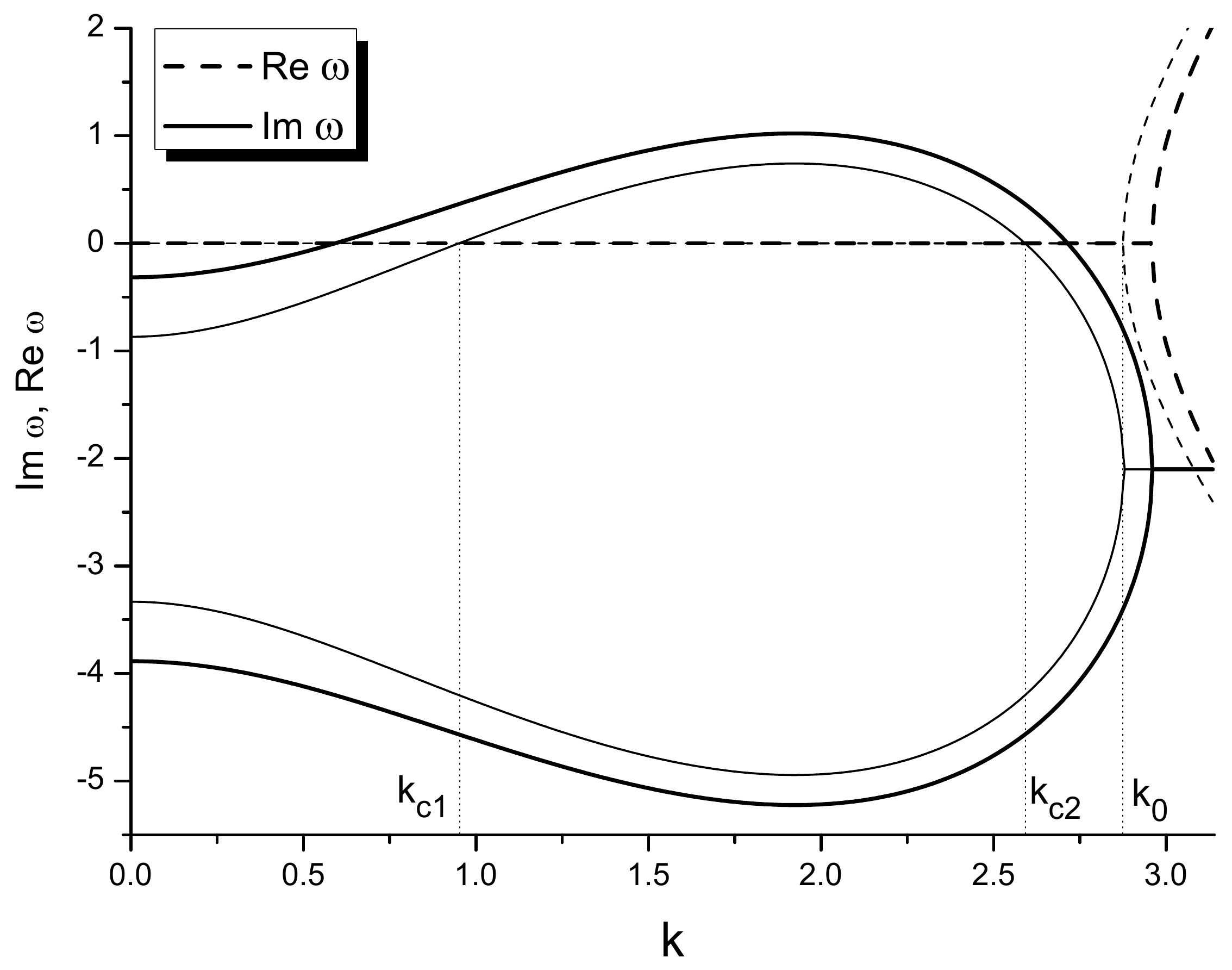} b)\includegraphics[width=80mm]{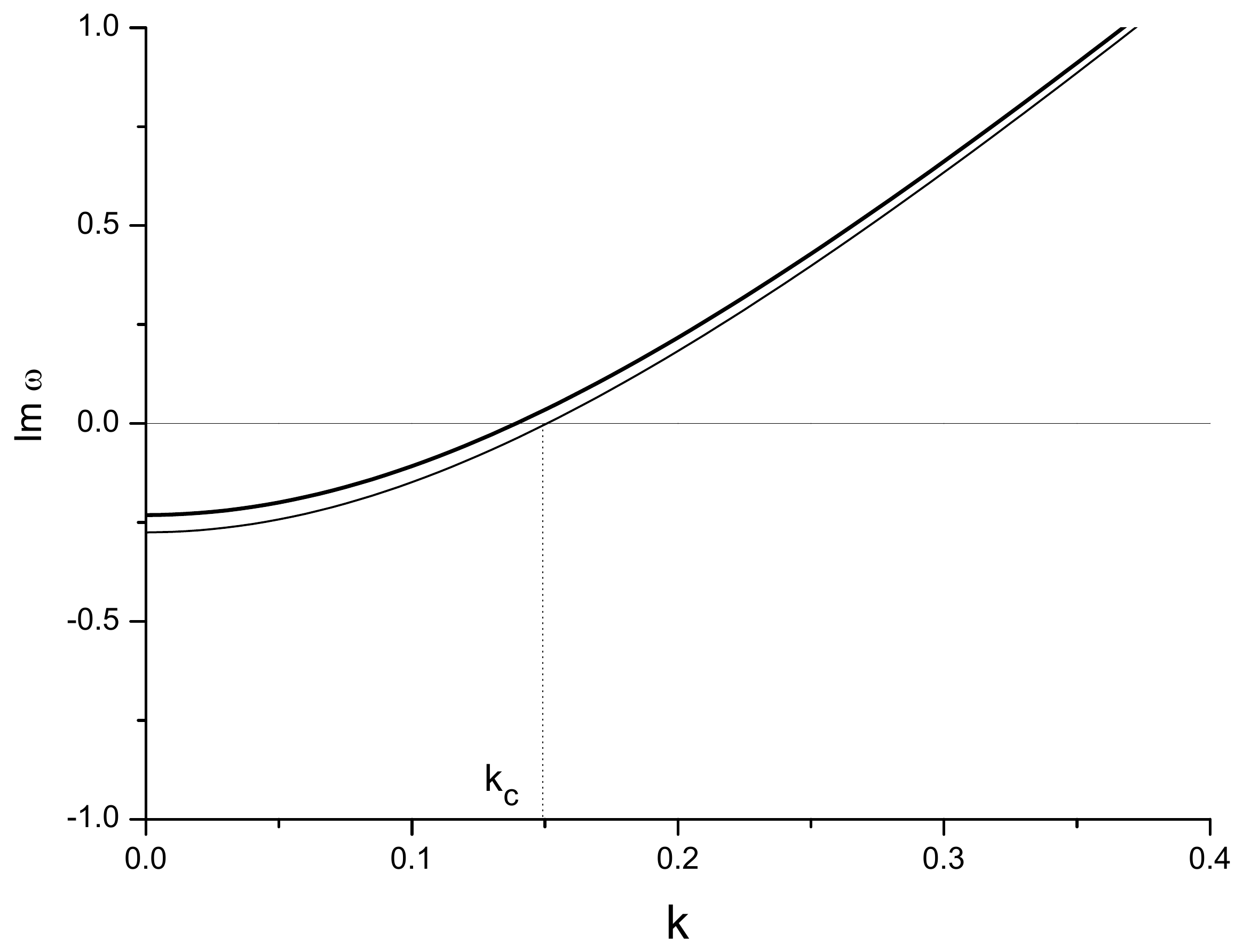}
\caption{Real and imaginary parts of the dispersion relation $\omega(k)$ for
deterministic and stochastic systems (thin and thick lines, respectively) at
$\varepsilon=2.5$, $D_0=1$, $\tau_J=0.3$: a) $\alpha=0.03323$, $x_0=0.035$
(thin and thick lines relate to $\sigma^2=0$ and $\sigma^2=0.2$, respectively);
b) $\alpha=0.4055$, $x_0=0.75$ (thin and thick lines relate to $\sigma^2=0$ and
$\sigma^2=0.1$, respectively)\label{det_x_small}}
\end{figure}

To make a quantitative analysis we consider firstly stability of the
deterministic system setting $\sigma^2=0$ in the vicinity of the unique minimum
position $x_0$ of $V(x)$. As far as $q_0$ increases with the lateral
interaction energy, next, we take small values for $\varepsilon$ and study two
cases related to small and large values for the adsorption rate. The
corresponding dependencies of $\Re\omega(k)$ and $\Im \omega(k)$ are shown in
Fig.\ref{det_x_small}.
From Fig.\ref{det_x_small}a obtained in the vicinity of diluted phase
characterized by $x_0\to 0$ modes with wave-numbers $k>k_{0}$ manifest decaying
oscillations (here $\Re\omega_+(k)=-\Re\omega_-(k)\ne0$, $\Im\omega_+(k)<0$).
Inside both intervals $k\in[0,k_{c1}]$ and $k\in[k_{c2},k_{0}]$ the
corresponding modes are stable without oscillations (here $\Re\omega_\pm(k)=0$,
$\Im\omega_\pm(k)<0$). In the domain $k\in(k_{c1},k_{c2})$ unstable modes
characterized by $\Im\omega_+(k)>0$ and $\Re\omega_\pm(k)=0$ appear. In the
vicinity of the dense phase with $x_0\to 1$ (see Fig.\ref{det_x_small}b) there
are no oscillatory solutions due to $\Re\omega_\pm(k)=0$, but unstable modes
can be realized. Here at small wave-numbers one has $\Im\omega_\pm(k<k_c)<0$,
whereas at $k>k_c$ all modes are unstable in the linear regime with
$\Im\omega_+(k)>0$\footnote{In the simplest case of $\varkappa=0$ in expansion
(\ref{expansionU}) one can find that in the vicinity of the diluted phase with
above set of parameters $\varepsilon$ and $\alpha$ all modes manifest decaying
oscillations, there are no unstable modes, here $\Re\omega_\pm(k)<0$,
$\Im\omega_+(k)>0$. In the vicinity of the dense phase one has
$\Re\omega_\pm<0$, whereas $\Im\omega_+(k)>0$ at $k>k_0$}. In the stochastic
case ($\sigma^2\ne0$) in the vicinity of the diluted phase the noise action
results in a decrease of the number of oscillating modes shifting $k_{0}$
toward $\pi$; the interval for unstable modes $[k_{c1},k_{c2}]$ enlarges (see
thick lines in Fig.\ref{det_x_small}a). In the vicinity of the dense phase the
noise shifts position of the critical wave-number $k_c$ toward small $k$
increasing the number of unstable modes (see thick lines in
Fig.\ref{det_x_small}b).

\begin{figure}
\centering
 \includegraphics[width=80mm]{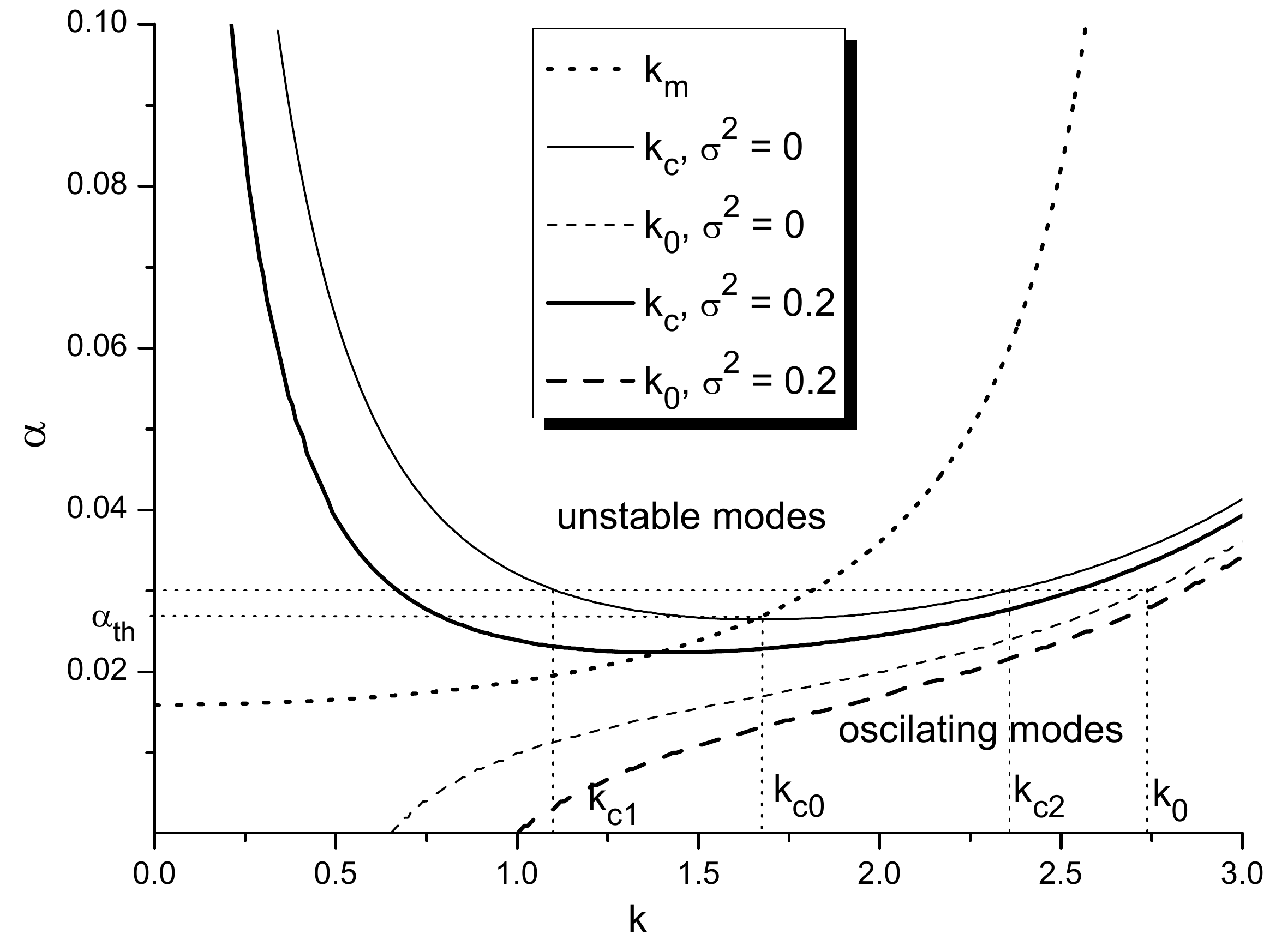}
\caption{Dependencies $\alpha(k)$ at $\sigma^2=0$ (thin lines) and
$\sigma^2=0.2$ (thick lines). Other parameters are: $\varepsilon=2.5$, $D_0=1$,
$\tau_J=0.3$. Domains of unstable modes and stable oscillating modes are
bounded by solid and dashed lines, respectively. \label{alpha(k)}}
\end{figure}

In Figure \ref{alpha(k)} we plot stability diagram $\alpha(k))$ at fixed
$\varepsilon$ for both deterministic and stochastic cases.
Here above solid lines (thin for the noiseless case and thick for the
stochastic one) the corresponding modes with $k\in[k_{c1}, k_{c2}]$ are
unstable. Below dashed lines the stable modes manifest oscillating behavior.
The first unstable mode $k_{c0}$ is realized when equation $\Im\omega(k)_+=0$
has unique nonzero solution $k_{c0}$. At small lateral interaction strength of
the adsorbate and small adsroption rate only stable modes without oscillations
and stable oscillating modes can be formed. Above the threshold
$\alpha_{th}=\alpha(k_{c0})$ unstable modes appear leading to spatial
self-organization of the adsorbate. When the noise intensity increases above
pattern forming processes can be observed at lower value for the adsorption
rate; the corresponding patterns are characterized by large wave-lengths then
in the noiseless case. The thick dotted line denotes the dependence of the
wave-number $k_m$ related to a position of the maximum for $\Im\omega(k)_+$. At
$k>k_{c0}$ the corresponding value $k_m$ defines the most unstable mode giving
a period of the spatial structure. The most unstable mode at $k>k_{c0}$ is
characterized by
\begin{equation}
k_m^2=\frac{q_0^2}{3}\left(2-\sqrt{1+\frac{3}{q_0^2G(x_0)}}\right).
\end{equation}
It follows that $k_m$ depends only on the adsorption rate and the lateral
interaction energy. The noise action leads to a change in the position of the
point $k_{c0}$ lying in the line $\alpha(k_m)$.

\begin{figure}
\centering
 \includegraphics[width=80mm]{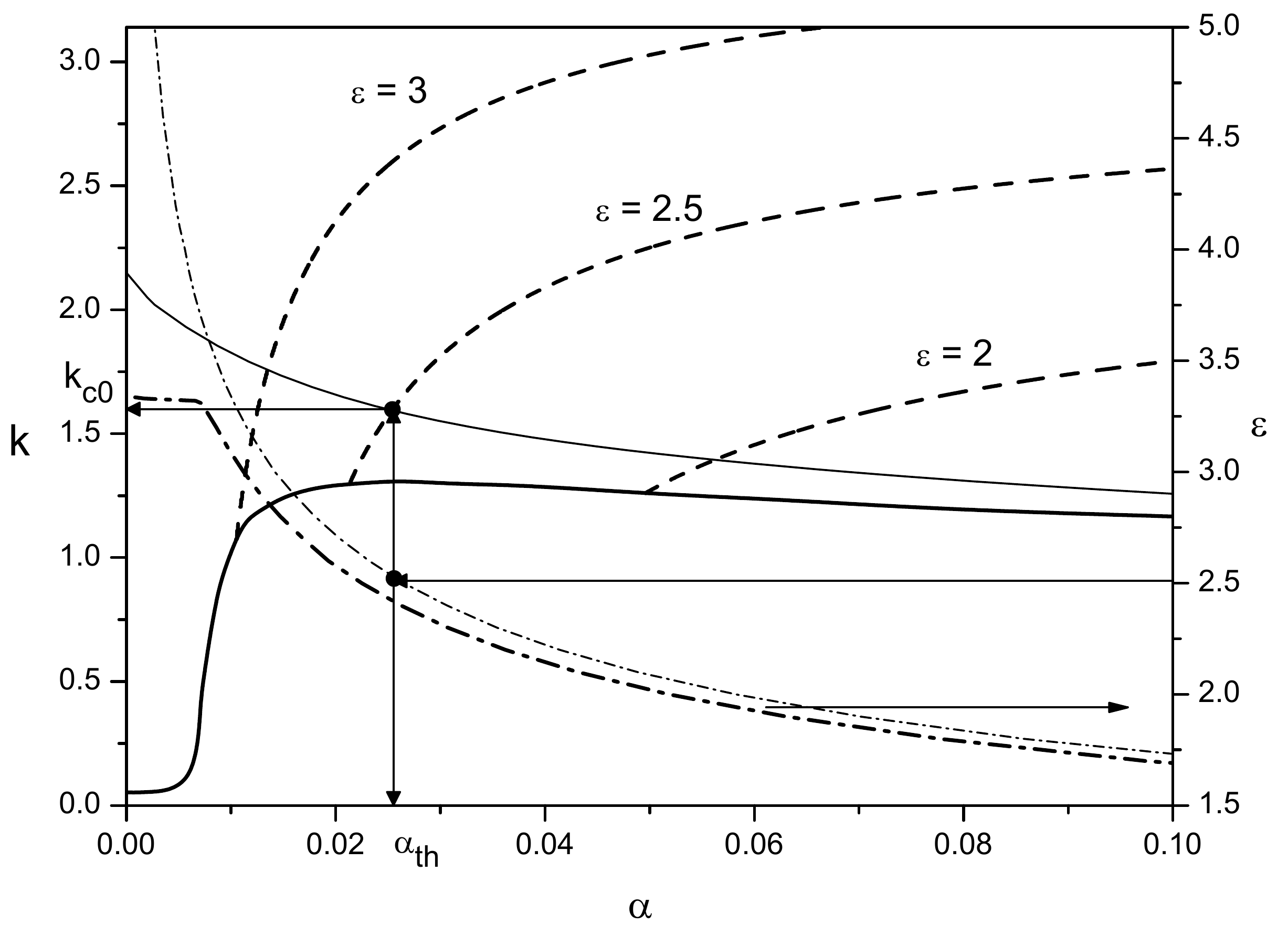}
\caption{Dependencies $k_m(\alpha)$ (see left ordinate axis) as dashed lines
starting at $k=k_{c0}$ shown for $\sigma^2=0.2$ and threshold values as
dependencies $\varepsilon(\alpha)$ (see right ordinate axis) at $\sigma^2=0$
and $\sigma^2=0.2$ (thin and thick solid lines, respectively). Other parameters
are: $D_0=1$, $\tau_J=0.3$. \label{km_alpha_epsilon}}
\end{figure}

Dependencies of the period of patterns at different values for lateral
interaction energy are shown in Fig.\ref{km_alpha_epsilon} as dashed lines
$k_m(\alpha)$ starting at $k=k_{c0}$ for $\sigma^2=0.2$. Here choosing a fixed
value for $\varepsilon$ from dependencies $\varepsilon(\alpha)$ one can find
both the threshold value $\alpha_{th}$ and the corresponding $k_{c0}$. Above
the obtained $k_{c0}$ the related period of patterns $\Lambda=2\pi/k_m$ can be
found at fixed system parameters. It follows that as the noise intensity
increases the unstable modes appear at lower values for both $\alpha$ and
$\varepsilon$. At elevated $\varepsilon$ the selected patterns are
characterized by small period.

As far as the first statistical moment illustrates oscillatory behavior one
should expect that processes of pattern selection can be observed in such
system. Considering properties of pattern selection we need to find dynamical
equation for the structure function $S(\mathbf{k},t)$ as a Fourier transform of
the two point correlation function $\langle\delta x(\mathbf{r},t)\delta
x(\mathbf{r}',t)\rangle$ and to study its oscillatory behavior at small times.
To that end we obtain the linearized evolution equation for the Fourier
components $\delta x_{\mathbf{k}}(t)$ and $\delta x_{-\mathbf{k}}(t)$ and
compute $S(\mathbf{k},t)=\langle\delta x_\mathbf{k}(t)\delta
x_{-\mathbf{k}}(t)\rangle$. Using Novikov's theorem, and holding only leading
terms in it, one gets a dynamical equation
\begin{equation}
\begin{split}
\tau_J\partial^2_{tt}S(\mathbf{k},t)&+\gamma(x_0)\partial_tS(\mathbf{k},t)\\
 &=2\left(f'(x_0)-\frac{\sigma^2\tau_J}{2}f'''(x_0)-D_0k^2(1-G(x_0)[q_0^2-k^2]^2)\right)S(\mathbf{k},t)+2\sigma^2\gamma(x_0).
\end{split}
\end{equation}
Its analytical solution can be found in the form $S(\mathbf{k},t)-S_0\propto
\exp(-i\varpi(\mathbf{k})t)$, where
\begin{equation}
\varpi(k)_\pm=-\frac{{\rm
i}\gamma(x_0)}{2\tau_J}\pm\sqrt{\frac{2\left(D_0k^2(1-G(x_0)[q_0^2-k^2]^2)-f'(x_0)\right)}{\tau_J}+\sigma^2f'''(x_0)-\frac{\gamma^2(x_0)}{4\tau^2_J}}.
\end{equation}
As in the previous case $\Im\varpi$ is responsible for stability of the system,
whereas $\Re\varpi$ relates to pattern selection processes. Unstable modes are
defined according to Eq.(\ref{kc}). The quantity $k_0$ is determined as
solution of the equation
\begin{equation}
k_0^2(1-G(x_0)[q_0^2-k_0^2]^2)=\frac{1}{D_0}\left[f'(x_0)-\frac{\sigma^2\tau_J}{2}f'''(x_0)+\frac{\gamma^2(x_0)}{8\tau_J}\right].
\end{equation}

\begin{figure}[!h]
\centering
 a)\includegraphics[width=50mm]{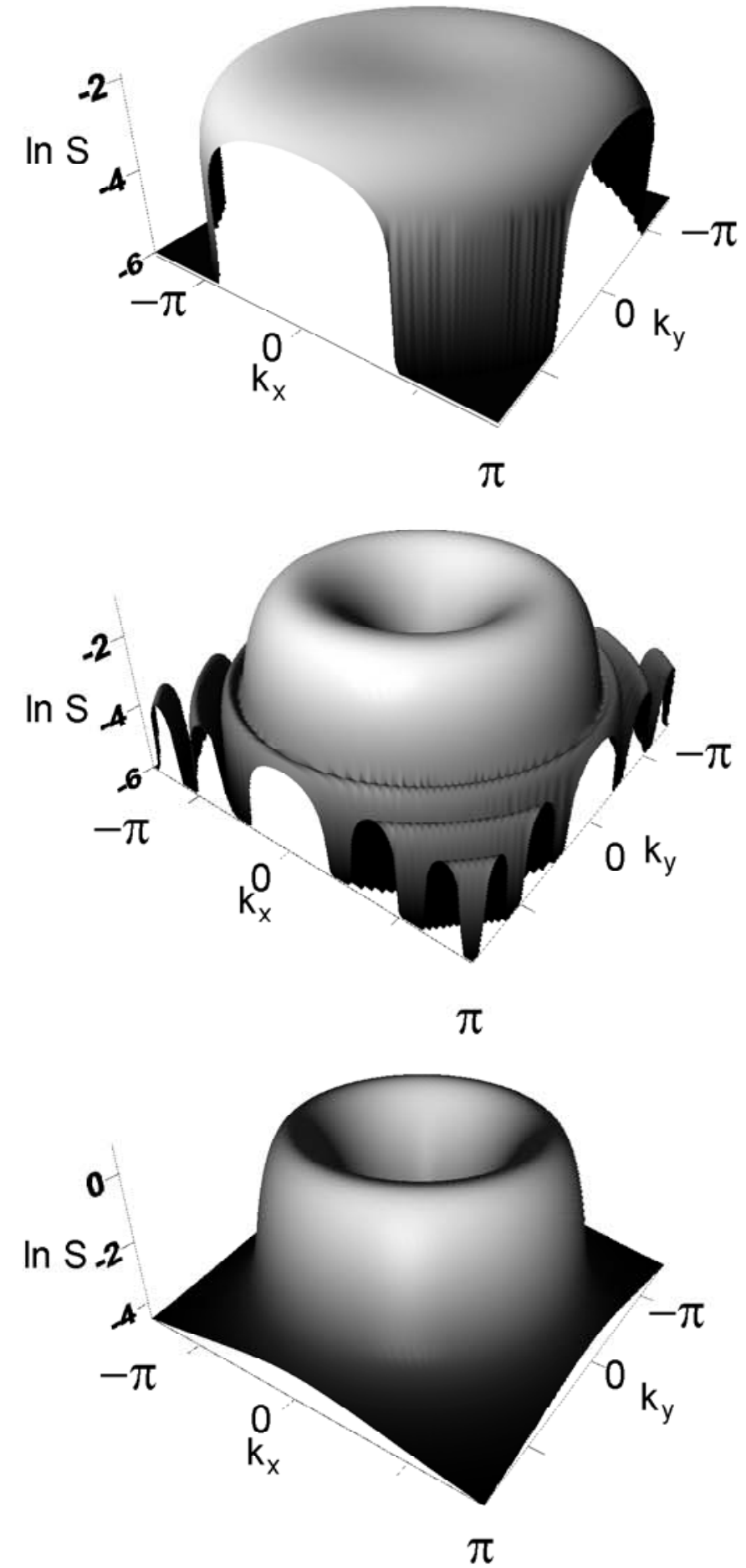}\ b)\includegraphics[width=80mm]{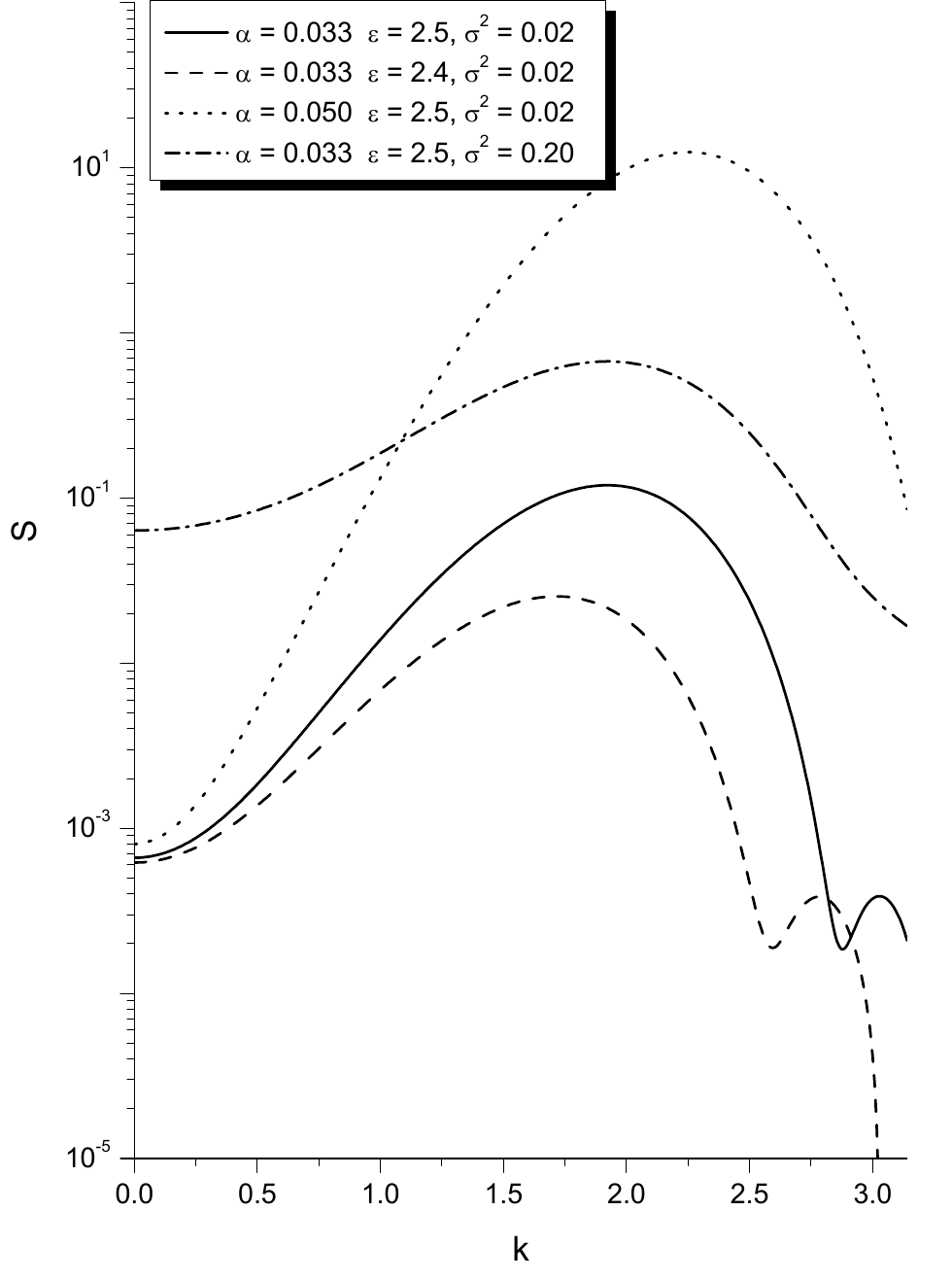}
\caption{(a) The structure function dynamics at early stages $t=1$, 2.4, 3 (the
time is increased from top to bottom) in the vicinity of the state $x_0=0.035$
at $\varepsilon=2.5$, $\alpha=0.03323$, $D_0=1$, $\sigma^2=0.02$. (b) The
structure function at $t=2.4$ and different values for $\varepsilon$, $\alpha$
and $\sigma^2$. \label{s(kt)}}
\end{figure}

An evolution of the structure function at early stages is shown in
Fig.\ref{s(kt)}a.
It is seen that during the system evolution the main peak related to
wave-number of most unstable mode is formed; its height is increased. At small
time interval except main unstable mode having largest peak in $S(k,t)$
additional peaks at $k>k_0$ correspond to selecting modes
\cite{PhysA1_2010,CEJP}. As far as these modes are stable during the system
evolution they disappear. Therefore, due to hyperbolic transport pattern
selection processes can be realized in the adsorption-desorption systems. In
Fig.\ref{s(kt)}b we plot the structure function at early stages at different
values for lateral interaction energy and noise intensity. As diagram in
Fig.\ref{km_alpha_epsilon} shows an increase in both $\varepsilon$ and $\alpha$
shifts the main peak position of $S(k)$ toward large values for the
wave-number. The noise action does not shift the main peak position. The peak
spreads due to fluctuation effects resulting in more diffuse interfaces between
two phases. Moreover, as the lateral interaction energy increases the pattern
selection processes occur faster than at small $\varepsilon$. The same is
observed for large adsorption rate and noise intensity.

\section{The effective Fokker-Planck equation and its stationary solution }

To consider statistical properties of the system at large time scales or in
stationary case one can use the corresponding Fokker-Planck equation. To obtain
it we exploit the standard technique and use the equivalent system of two
differential equations (\ref{eqX},\ref{Jtot}) written for both generalized
coordinate $x_i$ and momentum $p_i$ on a grid of the mesh size $\ell$:
\begin{equation}\label{eqPX}
\begin{split}
&\tau_J\dot x_i=p_i, \\ &\dot p_i+\frac{\gamma(x)}{\tau_J}p_i=
\varphi(x_i;\nabla_{ij}/\ell)+g(x_i)\zeta_i(t),
\end{split}
\end{equation}
here $i$ enumerates cells. Then, the corresponding probability density function
is given by the average of the density functional
$\overline{\rho(\{x_i\},\{p_i\},t)}$ over noise:
$P(\{x_i\},\{p_i\},t)\equiv\langle\overline{\rho(\{x_i\},\{p_i\},t)}\rangle$,
where $\overline{\cdots}$ means average over initial conditions. To obtain an
equation for the macroscopic density functional $P$ we exploit the conventional
device and proceed from the continuity equation:
\begin{equation}\label{ceq}
\frac{\partial \overline\rho}{\partial t}+\sum_i\left[\frac{\partial }{\partial
x_i}\dot x_i +\frac{\partial }{\partial p_i}\dot p_i\right]\overline\rho=0.
\end{equation}
Inserting $\dot x_i$ and $\dot p_i$ from Eq.~(\ref{eqPX}), we obtain
\begin{equation}\label{5_ce7}
\frac{\partial\overline\rho}{\partial t}= \left(\widehat{\mathcal{L}}+
\widehat{\Xi}\right)\overline\rho,
\end{equation}
where the operators $\widehat{\mathcal{L}}=\sum_i\widehat{\mathcal{L}}_i$ and
$\widehat{\Xi}=\sum_i\widehat{\mathcal{N}}_{i}\zeta_i$ are defined as follows:
\begin{equation}\begin{split}
&\widehat{\mathcal{L}}_i\equiv-\frac{p_i}{\tau_D}\frac{\partial}{\partial x_i}
+\frac{\partial}{\partial p_i}\left(-\varphi_i+\frac{\gamma_i}{\tau_D}p_i
\right),\\ &\widehat{\mathcal{N}}_i\equiv -g_{i} \frac{\partial}{\partial p_i}.
\end{split}
\end{equation}

Within the interaction representation, the density functional reads
$\wp=e^{-\widehat{\mathcal{L}}t}\overline\rho$, that allows us to rewrite
Eq.~(\ref{5_ce7}) as
\begin{equation}
\frac{\partial}{\partial t}\wp= \widehat{\mathcal{R}}\wp,\quad
\widehat{\mathcal{R}}=e^{-\widehat{\mathcal{L}}t}\widehat{\Xi}
e^{\widehat{\mathcal{L}}t}.\label{5_b}
\end{equation}
The well-known cumulant expansion method serves as a standard and effective
device to solve such a stochastic equation \cite{VanKampen}. Neglecting terms
of the order $O(\widehat{\mathcal{R}}^3)$, we get the kinetic equation for the
averaged quantity $\langle\wp(t)\rangle$ in the form
\begin{equation}\label{5_ce12}
\frac{\partial}{\partial t} \langle\wp\rangle(t)=\left[
\int\limits_0^t\langle\widehat{\mathcal{R}}(t)
\widehat{\mathcal{R}}(t')\rangle\D t'\right]\langle\wp\rangle(t).
\end{equation}
Integrating the right hand side, one gets
\begin{equation}
\begin{split}
\int\limits_0^t\langle\widehat{\mathcal{R}}(t) \widehat{\mathcal{R}}(t')\rangle
{\rm d}t'&=\int_0^t {\rm d}\tau \left<\sum_i\widehat{\mathcal{N}}_{i}\zeta_i
e^{\widehat{\mathcal{L}}\tau}\sum_j\widehat{\mathcal{N}}_{j}\zeta_j
e^{-\widehat{\mathcal{L}}\tau}\right>\\
 &=\sigma^2\int_0^t {\rm d}\tau \sum_{i}C(\tau)\widehat{\mathcal{N}}_{i}
e^{\widehat{\mathcal{L}}\tau}\widehat{\mathcal{N}}_{i}
e^{-\widehat{\mathcal{L}}\tau}.
\end{split}
\end{equation}
 Within the original representation, the equation for the
probability density reads
\begin{equation}
\frac{\partial}{\partial t}P=\left\{\widehat{\mathcal{L}}+
\sigma^2\int\limits_0^t\D\tau
\sum_{i}C(\tau)\left[\widehat{\mathcal{N}_i}\left(
e^{\widehat{\mathcal{L}}\tau}\widehat{\mathcal{N}_i}
e^{-\widehat{\mathcal{L}}\tau} \right)\right]\right\}P.
\end{equation}
Because the physical time is much larger than the correlation scale, one can
replace the upper limit of the integration with $\infty$.

To proceed we use the procedure proposed in Ref.\cite{PRE93,PRE2005,EPJB2003}
to obtain the effective Fokker-Planck equation for the stochastic model. Then,
expanding exponents, we obtain for the perturbation expansion
\begin{equation}
\frac{\partial P}{\partial
t}=\left(\widehat{\mathcal{L}}+\widehat{\mathcal{C}}\right)P \label{FP}
\end{equation}
where collision operator $\mathcal{C}$ is defined as follows:
\begin{equation}\label{Cop}
\widehat{\mathcal{C}}=\sum\limits_{n=0}^\infty\mathcal{M}^{(n)}\sum_{i}
\widehat{\mathcal{N}_i} \widehat{\mathcal{L}_i}^{(n)},\quad
\widehat{\mathcal{L}_i}^{(0)} =\widehat {\mathcal{N}_i},
\end{equation}
here
$\widehat{\mathcal{L}_j}^{(n)}=[\widehat{\mathcal{L}},\widehat{\mathcal{L}_j}^{(n-1)}]$
($[A,B]$ is the commutator); moments of the noise temporal correlation function
$C(\tau)$ are: $\mathcal{M}^{(n)}=(\sigma^2/{n!})\int_0^\infty
\tau^nC(\tau){\rm d}\tau$.

To perform the following calculations we shall restrict ourselves to
considering overdamped systems where the variation scales $t_s$, $\ell$, $x_s$,
$v_s$, $\gamma_s$, $f_s$, $D_s$, and $g_s$ of the time $t$, the coordinate $\bf
r$, the quantity $x$, the velocity $v\equiv p/\tau_J$, the damping coefficient
$\gamma(x)$, the force $f(x)$ and the noise amplitude $g(x)$, respectively,
obey the following conditions:
\begin{equation}\label{scale}
\begin{split}
\frac{v_s t_s}{x_s}\equiv\epsilon^{-1}\gg 1,\quad \frac{\gamma_s
t_s}{\tau_J}\equiv\epsilon^{-2}\ggg 1,\\ \frac{f_s t_s}{v_s \tau_J}=\frac{g_s
t_s}{v_s \tau_J}\equiv\epsilon^{-1}\gg 1, \quad\frac{D_s x_s t_s}{\tau_J
v_s\ell^2}=\epsilon^{-1}\gg 1.
\end{split}
\end{equation}
These conditions means a hierarchy of the damping and the
deterministic/stochastic forces characterized by relations
\begin{equation}\label{scale1}
\frac{f_s}{\gamma_s v_s}=\frac{g_s}{\gamma_s v_s}\equiv\epsilon\ll
1,\quad\frac{D_s x_s}{\gamma_s v_s\ell^2}\equiv\epsilon\ll 1.
\end{equation}

As a result, the dimensionless system of equations (\ref{eqPX}) takes the form
\begin{equation}\label{eq22}
\begin{split}
\frac{\partial x_i}{\partial t}=&~\epsilon^{-1}v_i,\\ \frac{\partial
v_i}{\partial t}=&-\epsilon^{-2}\gamma_i v_i+\epsilon^{-1}\left[\varphi_i
 +g_{i}\zeta_{i}(t)\right].
\end{split}
\end{equation}
Respectively, the Fokker--Planck equation (\ref{FP}) reads
\begin{equation}\label{Ptheta}
\left(\frac{\partial }{\partial
t}-\widehat{\mathcal{L}}\right)P=\epsilon^{-2}\widehat{\mathcal{C}}P
\end{equation}
where the operator
\begin{equation}
\widehat{\mathcal{L}}\equiv\epsilon^{-1}\widehat{\mathcal{L}}_1
+\epsilon^{-2}\widehat{\mathcal{L}}_2
\end{equation}
has the components
\begin{equation}
\widehat{\mathcal{L}}_1\equiv \sum_i-v_i\frac{\partial}{\partial x_i}
-\varphi_i \frac{\partial}{\partial v_i},\quad
\widehat{\mathcal{L}}_2\equiv\sum_i\gamma_i~\frac{\partial}{\partial v_i}v_i.
\end{equation}
The collision operator is defined through the expressions type of (\ref{Cop}):
\begin{equation}
\begin{split}
\epsilon^{-2}\widehat{\mathcal{C}}=\sum\limits_{n=0}^\infty
{\widehat{\mathcal{C}}^{(n)},\quad}\widehat{\mathcal{{C}}}^{(n)} =\sum_{i}
\mathcal{M}^{(n)}
\left(\widehat{\mathcal{N}}_i\widehat{\mathcal{L}}_i^{(n)}\right),\\
\widehat{\mathcal{L}}_i^{(0)} =\epsilon^{-2}\widehat {\mathcal{N}}_i,\quad
\widehat{\mathcal{N}}_i\equiv-g_{i}\frac{\partial}{\partial v_i}.
\end{split}
\end{equation}
After suppressing the factor $\epsilon^{-2}$, the collision operator written
with accuracy up to the first order in $\epsilon\ll 1$ takes on the explicit
form up to second order in $\epsilon$:
\begin{equation}
\begin{split}
 \widehat{\mathcal{C}}=&\sum_{i}\left\{\left(\mathcal{M}^{(0)}-\gamma_i
\mathcal{M}^{(1)}\right) g_{i}^2\frac{\partial^2}{\partial v_i\partial
v_i}\right.\\&\left.+\epsilon \mathcal{M}^{(1)}g_{
i}^2\left[\frac{\partial^2}{\partial x_i\partial
 v_i}-\frac{1}{g_{
i}}\left(\frac{\partial g_{i}}{\partial
x_i}\right)\left(\frac{\partial}{\partial v_i} +v_i\frac{\partial^2}{\partial
 v_i\partial
 v_i}\right)\right]\right\}.
\end{split}
\end{equation}

Note that in further consideration we are interested in the behavior of the
particular distribution ${\mathcal P}(\{x_i\},t)$, not the total one,
$P(\{x_i\},\{v_i\},t)$. The reduced distribution can be obtained according to
the moments introduced as follows
\begin{equation}\label{5_one_site}
{\mathcal P}^{(n)}(\{x_i\},t)\equiv\int P(\{x_i\},\{v_i\},t)
\prod_i\left[v_i^n{\rm d}v_i\right],
\end{equation}
where the integration is provided over set~$\{v_i\}$. Then, performing
corresponding manipulations with Eq.~(\ref{FP}), we arrive at the recursive
relations for the moments ${\mathcal P}^{(n)}(\{x_i\},t)$ \cite{PRE2005}:
\begin{equation}\label{Pmoment}
\begin{split}
\epsilon^2\frac{\partial{\mathcal P}^{(n)}}{\partial t}+&n\gamma_i{\mathcal
P}^{(n)}+\epsilon\left[\frac{\partial{\mathcal P}^{(n+1)}}{\partial
x_i}-n\varphi_i{\mathcal P}^{(n-1)}\right]\\
&=n(n-1)\left(\mathcal{M}^{(0)}-\gamma_i \mathcal{M}^{(1)}
\right)g_{i}^2{\mathcal P}^{(n-2)}-\epsilon n
\mathcal{M}^{(1)}\left[g_{i}^2\frac{\partial{\mathcal P}^{(n-1)}}{\partial
x_i}+n g_{i}\left(\frac{\partial g_{i}}{\partial x_i}\right){\mathcal
P}^{(n-1)}\right].
\end{split}
\end{equation}
Here ${\mathcal {P}}\equiv {\mathcal P}^{(0)}(\{x_i\},t)$, the first moment
${\mathcal {P}}^{(1)}(\{x_i\},t)$ can be considered as a flux of the
probability density ${\mathcal{P}}$, i.e. ${\mathcal
P}^{(1)}\equiv\mathcal{J}$. Indeed, taking the zeroth moment, we obtain the
expected continuity equation
\begin{equation}\label{CEF}
\partial_t \mathcal{P}=-\frac{1}{\epsilon}\sum_i\frac{\partial}{\partial
x_i}\mathcal{J}.
\end{equation}
The first moment calculation leads to
\begin{equation}
\epsilon^2\partial_t\mathcal{J}=-{\sum_i\gamma_i}\mathcal{J}+\epsilon\sum_i\left[\varphi_i\mathcal{P}
-\frac{\partial \mathcal{P}^{(2)}}{\partial x_i} -{\mathcal{M}^{(1)}}
\left\{g^2_i\frac{\partial \mathcal{P}}{\partial x_i}+ g_i\left(\frac{\partial
g_i}{\partial x_i} \right)\mathcal{P}\right\}\right].
\end{equation}
To evaluate the second moment $\mathcal{P}^{(2)}$ we put $n=2$ and take into
account only terms of zeroth order in $\epsilon$:
\begin{equation}
\mathcal{P}^{(2)}=\left(\frac{\mathcal{M}^{(0)}}{\gamma_i}-\mathcal{M}^{(1)}\right)
 g_i^2\mathcal{P}.
\end{equation}

Combining all obtained terms we arrive at the system of two equations of the
form
\begin{equation}
\begin{split}
 &\partial_t \mathcal{P}=-\frac{1}{\epsilon}\sum_i\frac{\partial}{\partial
x_i}\mathcal{J},\\
&\epsilon^2\partial_t\mathcal{J}=-{\sum_i\gamma_i}\mathcal{J}
-\epsilon\sum_i\left[-\left(\varphi_i+ \frac{{\mathcal{M}^{(1)}}}{2}
\left(\frac{\partial \gamma_i}{\partial x_i}
\right)\right)\mathcal{P}+{\mathcal{M}^{(0)}}\frac{\partial}{\partial
x_i}\mathcal{P}\right].
\end{split}
\end{equation}
In the stationary case with no probability density current one gets the
solution of the quasi-Gibbs form
\begin{equation}
\mathcal{P}[x]\propto
\exp\left(-\frac{\mathcal{U}_{ef}[x]}{\mathcal{M}^{(0)}}\right),
\end{equation}
where the effective energy functional is
\begin{equation}\label{eqUx}
\mathcal{U}_{ef}[x]=-\int{\rm d}{\mathbf{r}}\left(\int{\rm d}x'
\varphi(x';\nabla)+\frac{\mathcal{M}^{(1)}}{2}\gamma(x)\right).
\end{equation}
When we know the effective functional $\mathcal{U}_{ef}[x]$ or its general
construction in quadratures it can be used to describe possible phase
transitions in the system or to analyze possibility of pattern formation.

\section{Noise-induced transitions}

\begin{figure*}
\centering
 a)\includegraphics[width=80mm]{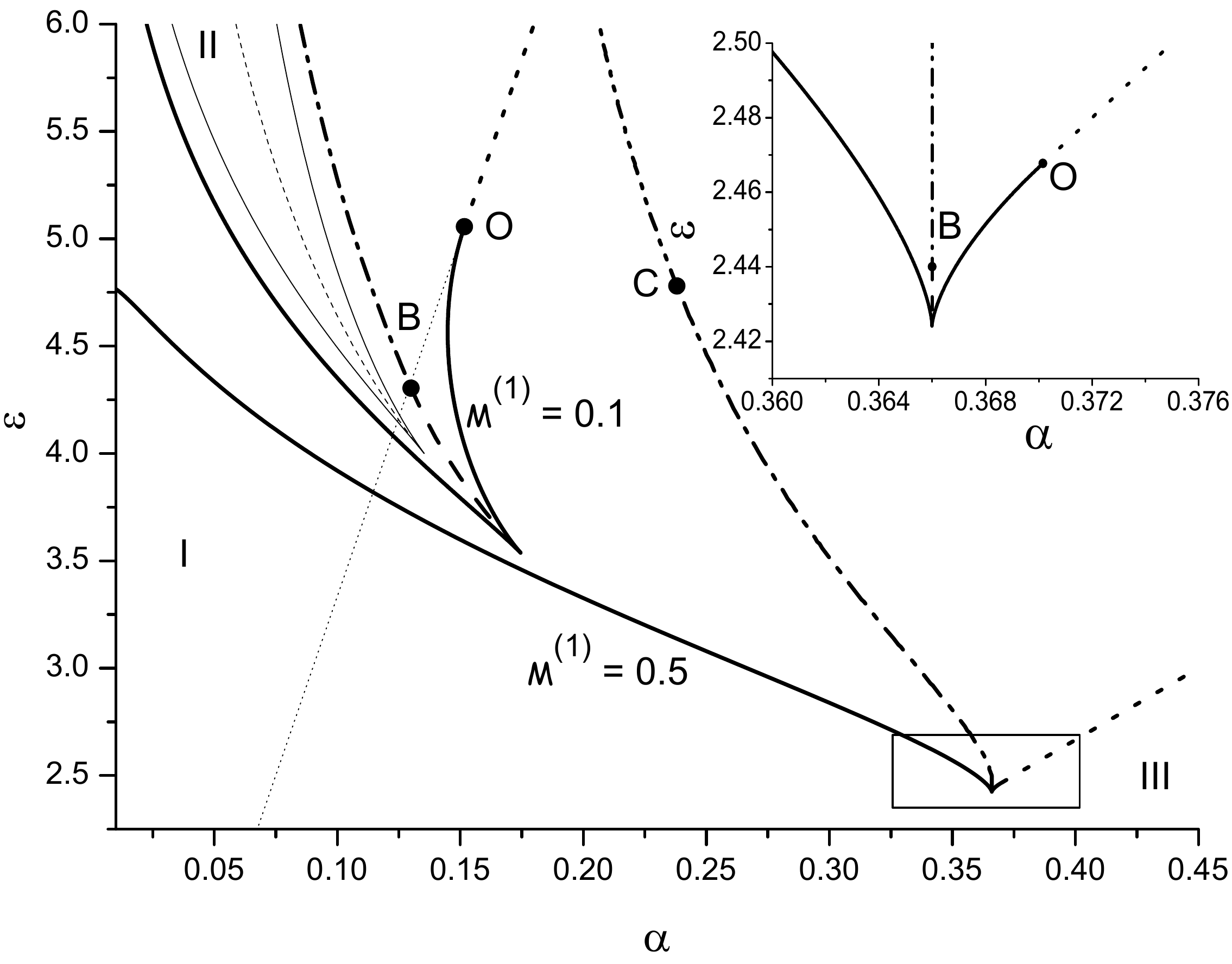}
 b)\includegraphics[width=80mm]{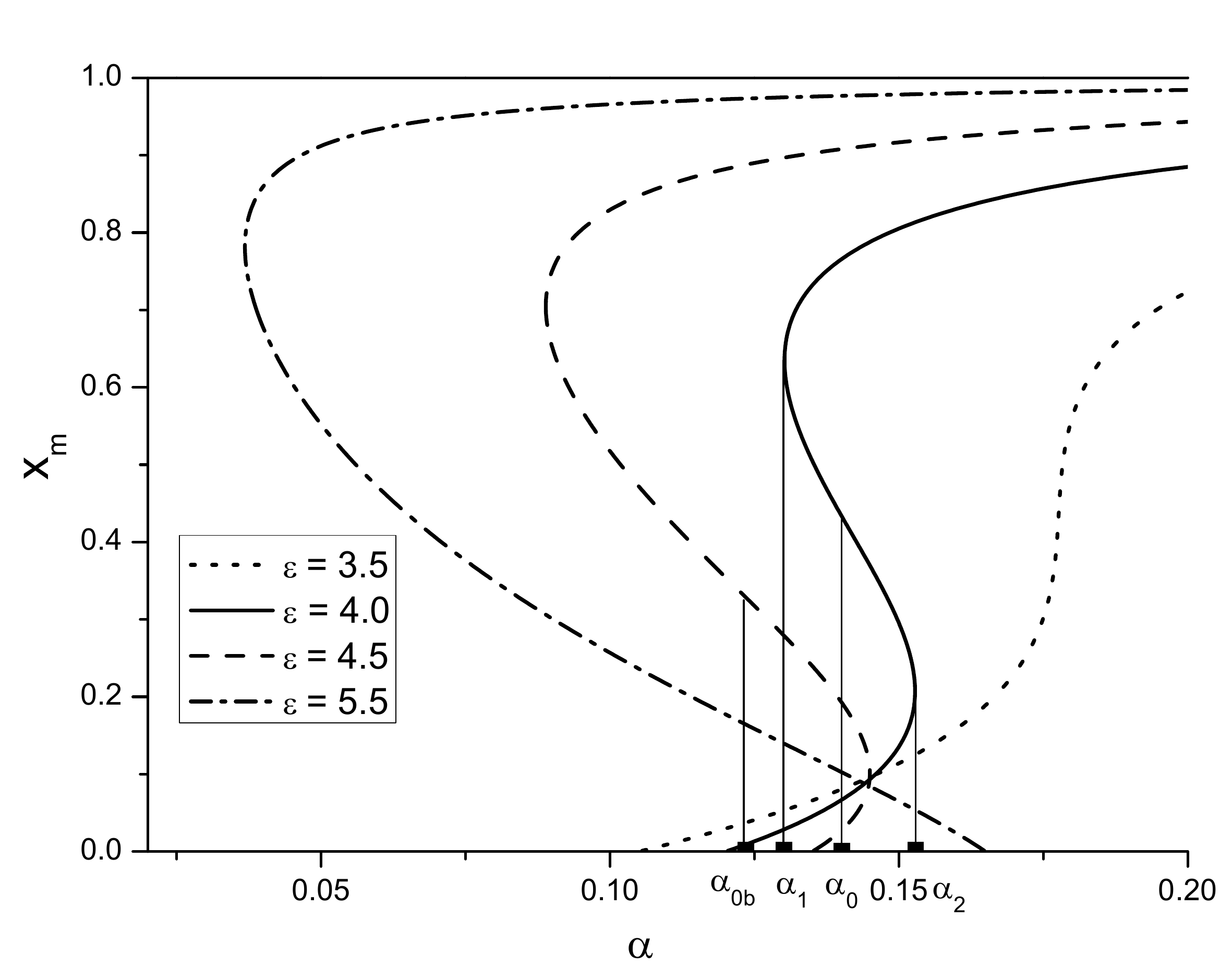}
 \caption{Phase and
bifurcation diagrams for non-equilibrium transitions at $\tau_J=0.3$: a) phase
diagram in the plane $(\alpha,\varepsilon)$ at pure deterministic case (thin
lines) and $\mathcal{M}^{(1)}=0.1$, 0.5 (thick lines);
 b) most probably stationary values $x_m$ versus $\alpha$ at different $\varepsilon$ and  $\mathcal{M}^{(1)}=0.1$.
 Domains $I$  and $III$  correspond to diluted  and dense states, respectively; in the domain $II$ both states are possible.
  \label{nit}}
\end{figure*}

Let us consider non-equilibrium (noise-induced) transitions in a homogeneous
system, setting $D_0=0$. The analysis of the effective potential
$U_{ef}(x)=-\int{\rm d} x'f(x')-\frac{c_0\mathcal{M}^{(1)}}{2}\gamma(x)$
results in noise-induced scenario for modality change of the stationary
distribution function $P(x)$. The corresponding diagrams for noise-induced
transitions are shown in Fig.\ref{nit}.
Dashed lines in Fig.\ref{nit}a correspond to coexistence of two phases:
equivalence of two minima of the function $V(x)=-\int{\rm d}x'f(x')$ (thin
line) and $U_{ef}(x)$ (thick line). The dash-dotted lines starting at the point
$B$ relate to the system parameters where the minimum of $U_{ef}$ at $x=x_m$
equals $U_{ef}(x=b_0)$, where $b_0=0$ is the lower natural boundary of the
diffusion process $x(t)$. From the mathematical viewpoint at such choice of the
system parameters one has a unique minimum at $x=x_m$, but due to existence of
boundary $b_0$ one has an ``effective minimum'' of $U_{ef}$ at $x=b_0$. This
minimum relates to the probability density of the adsorbate absence. Therefore,
the dash-dotted lines mean an equivalence of both the absence of the adsorbate
and the existence of the dense state with $x_m\ne 0$. The dotted line denotes
values for $\varepsilon$ and $\alpha$ at which $x_m$ takes zeroth values and
belongs to the boundary $b_0$. At large noise intensity
($\mathcal{M}^{(1)}=0.5$) the dash-dot-dot line starting at the point $C$
denotes equivalence of two ``effective'' minima of $U_{ef}$ at $x=b_0$ and at
$x=b_1$, where $b_1=1$ is the upper boundary of the diffusion process.
Therefore, the dash-dot-dot line corresponds to equivalence of both states
absence of the adsorbate and pure dense state. From Fig.\ref{nit}a it follows
that the colored noise results in a decrease of the critical values for the
system parameters $\alpha$ and $\varepsilon$ (cf. cases of white and colored
noise influence when $\mathcal{M}^{(1)}=0$ and $\mathcal{M}^{(1)}\ne 0$,
respectively). Therefore, if the system is subjected to the colored noise, then
non-equilibrium unimodal/bimodal transitions are possible at lower lateral
interaction energies $\varepsilon$ and adsorption rates $\alpha$. Bifurcation
diagram illustrating stationary values $x_m$ versus the adsorption rate
$\alpha$ is shown in Fig.\ref{nit}b. Here for the solid line quantities
$\alpha_1$ and $\alpha_2$ belong to spinodals (solid lines in Fig.\ref{nit}a);
$\alpha_0$ corresponds to the coexistence line; $\alpha_{0b}$ (see dashed line)
relates to the coexistence line of two states $x_m\ne 0$ and $x_m=b_0$.

\section{Analysis of stationary patterns}

The extreme of $\mathcal{U}_{ef}[x]$ relates to the stationary patterns
$x_{st}$ which can be computed setting the first variation of
$\mathcal{U}_{ef}[x]$ with respect to $x$ equal to zero, i.e. $\delta
\mathcal{U}_{ef}[x]=0$. From Eq.(\ref{eqUx}) it follows that stationary
patterns can be obtained as solutions of the problem
\begin{equation}\label{stpat}
f(x)+D_0\nabla\cdot[\nabla x-G(x)\nabla
\mathcal{L}_{SH}x]-\frac{c_0\tau_J\mathcal{M}^{(1)}}{2}f''(x)=0.
\end{equation}
In order to study stability of possible solution in the vicinity of the fixed
point $x_m$ related to extrema positions of the homogeneous distribution we
assume $x(r)-x_m\propto\exp({\rm i}\varkappa r)$, where $\varkappa$ can have
real and imaginary parts responsible for spatial modulations and stability of
the solution, respectively. After linearization of Eq.(\ref{stpat}) one has
\begin{equation}\label{eq_kapa}
D_0\varkappa^2(1-G(x_m)[q_0^2-\varkappa^2]^2)=f'(x_m)-\frac{\tau_J\tau_c\sigma^2}{2}f'''(x_m),
\end{equation}
where we put $c_0=1$, $\mathcal{M}^{(1)}=\tau_c\sigma^2$.

\begin{figure}[!t]
\centering 
 a) \includegraphics[width=80mm]{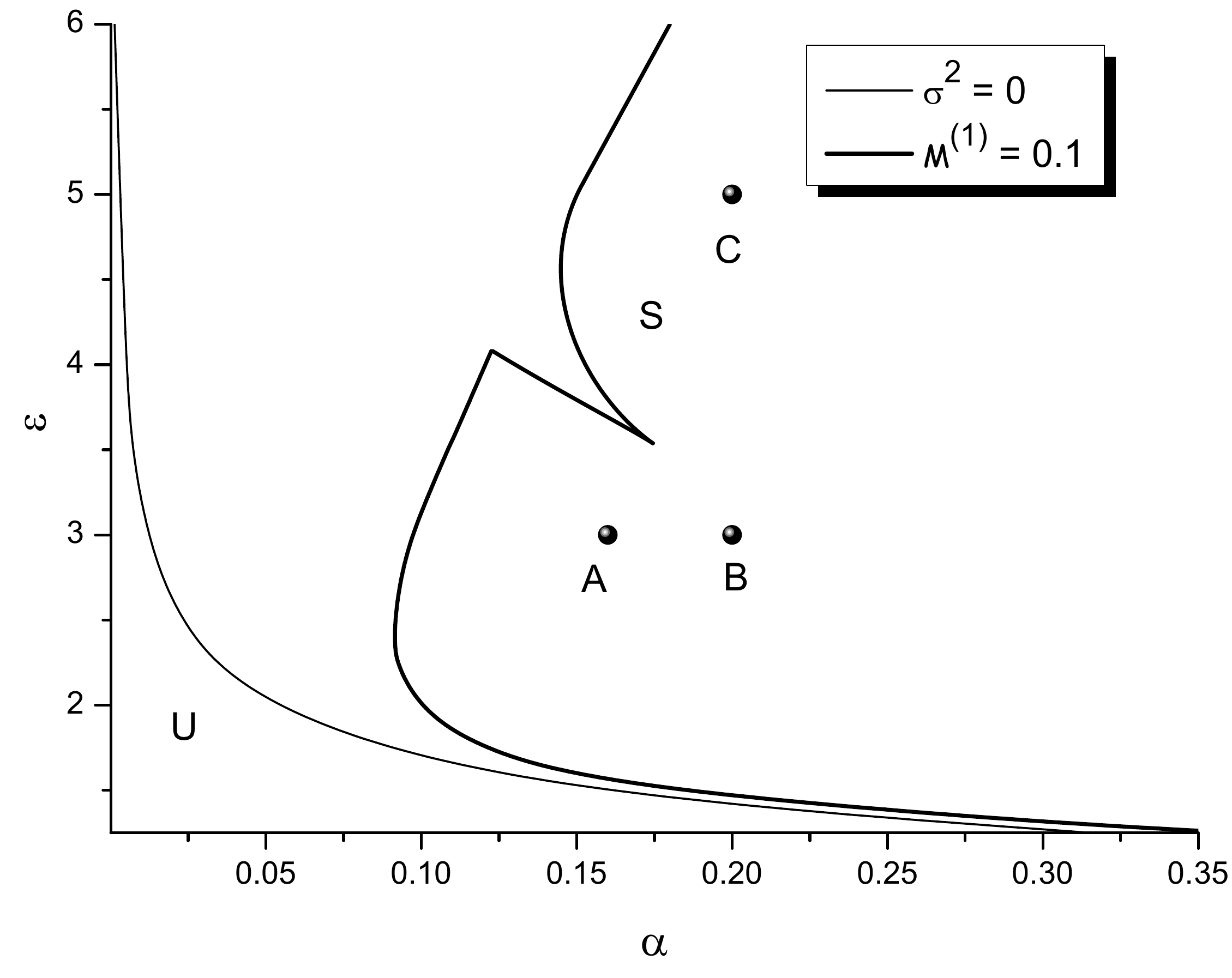}
 b) \includegraphics[width=80mm]{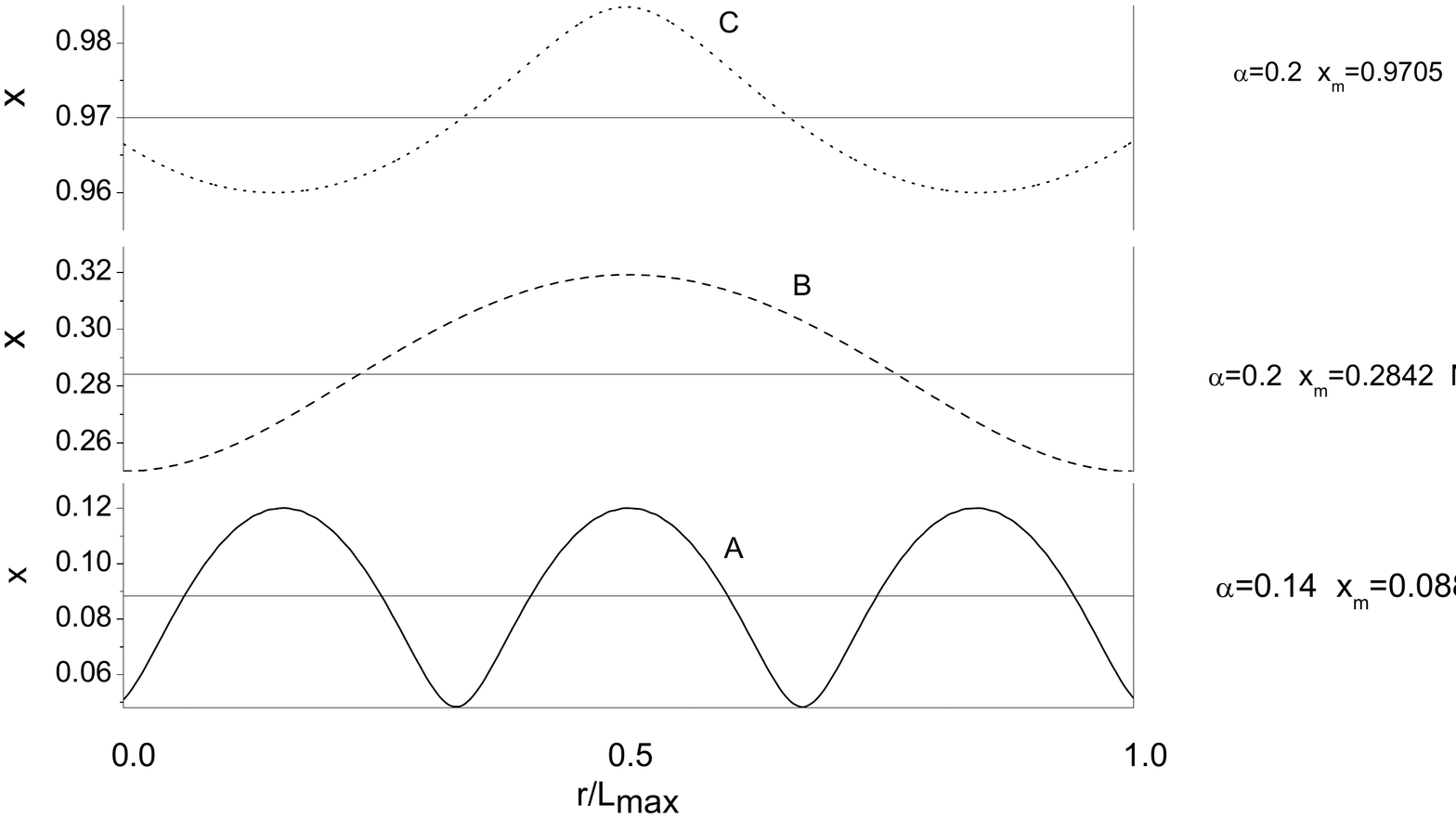}
 \caption{
(a) Stability diagram for stationary patterns: $U$ and $S$ denote domains of
unstable and stable patterns. (b) Profiles of stationary patterns related to
points $A(\alpha=0.14,\varepsilon=3.0)$, $B(\alpha=0.2,\varepsilon=3.0)$ and
$C(\alpha=0.2,\varepsilon=5.0)$ in plot (a) at $D_0=3.0$,
$\mathcal{M}^{(1)}=0.1$.\label{kapa}}
\end{figure}

According to obtained values for $\varkappa$ from Eq.(\ref{eq_kapa}) we have
computed a stability diagram shown in Fig.\ref{kapa}a for both stochastic
system with $\mathcal{M}^{(1)}=0.1$ and deterministic one. It is seen that in
the case of the noiseless system there is unique line of a hyperbolic form
dividing domains of unstable $(U)$ an stable $(S)$ patterns (see thin line). At
large $\varepsilon$ it tends to the axis $\alpha=0$. It follows that increasing
the lateral interaction energy or the adsorption rate independently one can
induce formation of stable stationary patterns. If colored fluctuations
($\mathcal{M}^{(1)}\ne 0$) are not negligibly small, then the picture of
stationary pattern formation can be changed crucially. Here the thick curve
dividing domains of stable and unstable stationary solutions at large
$\varepsilon$ tends to the dependence $\varepsilon(\alpha)$ denoting formation
of the first nonzero $x_m$ (line (OB) in Fig.\ref{nit}a). It follows that as
lateral interaction energy increases at fixed small values for adsorption rate
one can move from unstable domain into domain of stable stationary patterns. At
further increase in $\varepsilon$ we move again into domain of unstable
solutions. Therefore, a some kind of re-entrant pattern-forming transition is
realized. Starting from domain of unstable stationary patterns at elevated
values for $\alpha$ we move into the domain of stable stationary solutions.
Continuing an increase in $\varepsilon$ we get a re-entrantce, where unstable
stationary patterns can be observed. The corresponding part of the stability
diagram denoting re-entrant formation of unstable stationary patterns (a
``beak'') is a part of phase diagram describing noise-induced transitions (see
Fig.\ref{nit}a). At further increase in $\varepsilon$ one can move again from
domain of stable stationary patterns into domain of unstable stationary
patterns. Finally, at large $\alpha$ the re-entrant ordering (formation of
stable stationary patterns) can be observed. One dimensional stationary
patterns are shown in Fig.\ref{kapa}b where $L_{max}$ is the maximal period for
one from three patterns obtained at $\alpha$ and $\varepsilon$ related to
coordinates of points $A$, $B$ and $C$ in Fig.\ref{kapa}a. It is seen that at
small $\alpha$ one should expect a formation of islands of adsorbate. Such
situation is possible for deposition of molecules (polymers). At large $\alpha$
relevant for deposition of atoms on metal surface islands of diluted phase
(vacancy islands) can be organized.

\section{Quasi-stationary limit}
Let us consider evolution of islands of adsorbate or vacancies (inclusions)
having radius $R$ taking into account that the most probable state $x_m$ is
reached and the system fluctuates around it. We assume $(x_m-x_\infty)\le x_m$
where $x_\infty$ plays a role of thermodynamics concentration value, in other
words a supersaturation is small.

Let us introduce a quantity $y\equiv x_m-x_R$ measuring a deviation of the
inclusion form from a sphere of the radius $R$. Using the standard approach
\cite{LLX} one can assume the concentration in the vicinity of a spherical
inclusion as follows: $x_R=x_\infty(1+R_s/R)$, $R_s=2\nu\Omega/T$ is determined
trough the surface tension $\nu$, atomic volume $\Omega$ and temperature $T$.
Then, the diffusion flux $j(y)$ in the vicinity of the inclusion coincides with
velocity of its radius change: ${\rm d}R/{\rm d}t=j(y)$. Using the standard
formalism of the coalescence theory \cite{LLX}, we can replace the derivative
operator $\nabla$ in the flux
$j(y)=D_0\nabla(1-G(x_m)\nabla\mathcal{L}_{SH}(q_0,\nabla))y$ by $R^{-1}$. As a
result we obtain an evolution equation for the radius of inclusion in the form
\begin{equation}\label{rho}
\frac{{\rm d}R}{{\rm
d}t'}=\frac{1}{R}\left(1-\frac{G(x_m)}{R}\mathcal{L}_{SH}(q,R^{-1})\right)\left[\frac{1}{R_c}-\frac{1}{R}\right],
\end{equation}
where $t'\equiv tD_0\Sigma$, $\Sigma\equiv 2\nu\Omega x_\infty/T$,
$\Delta\equiv x_m-x_\infty\ge0$ plays the role of a supersaturation,
$R_c\equiv\Sigma/\Delta>0$ is the critical radius of the inclusion,
$\mathcal{L}_{SH}(q,R^{-1})\equiv\left(q^2+{R}^{-2}\right)^2$; next we drop
prime for the time variable. Equation (\ref{rho}) can be rewritten in the form
$\dot R=-{{\rm d}W}/{\rm d}R$, where a potential $W(R)$ can be used in a
further study.

\begin{figure}
\centering
 a)\includegraphics[width=80mm]{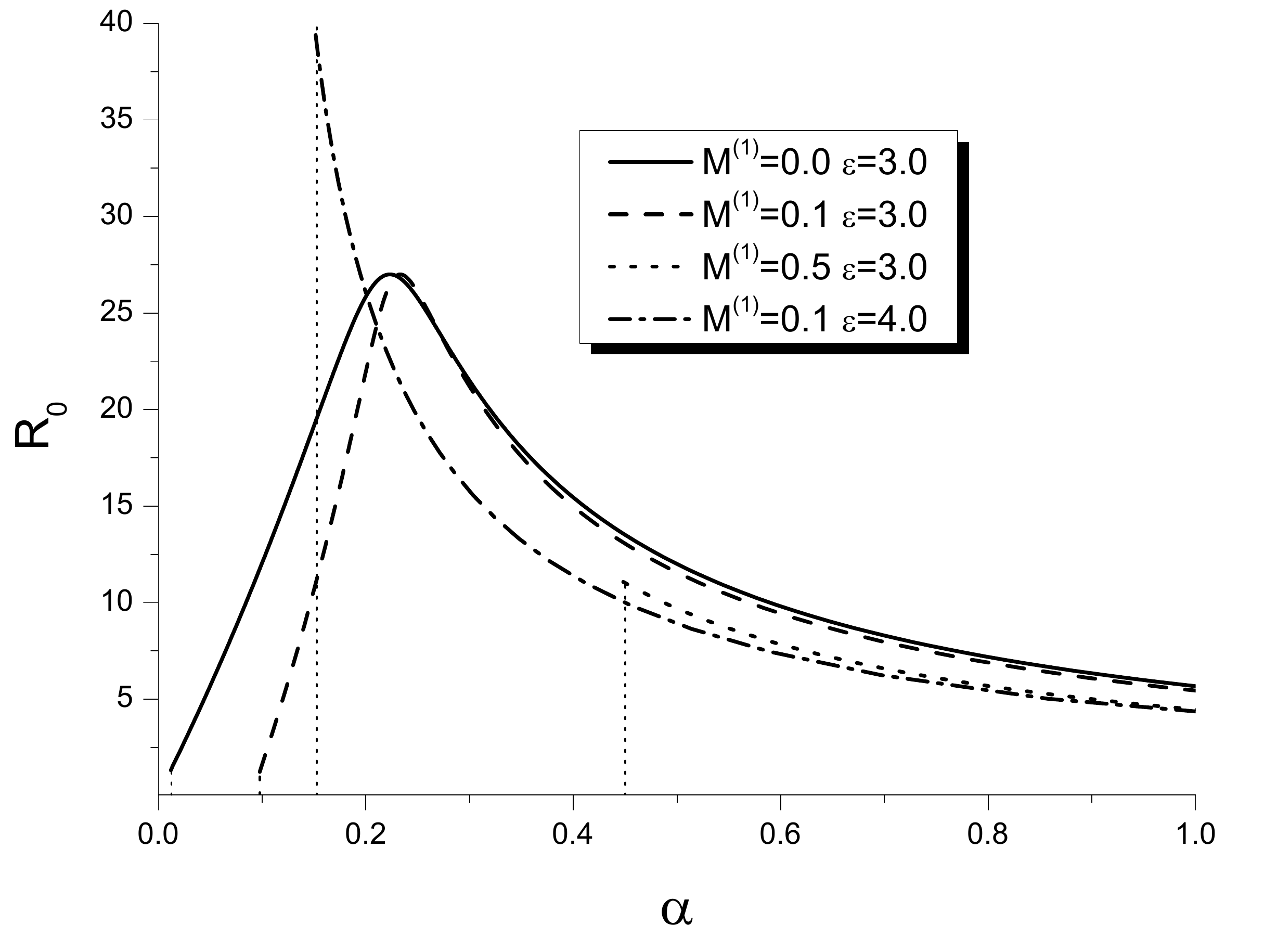}
 b)\includegraphics[width=80mm]{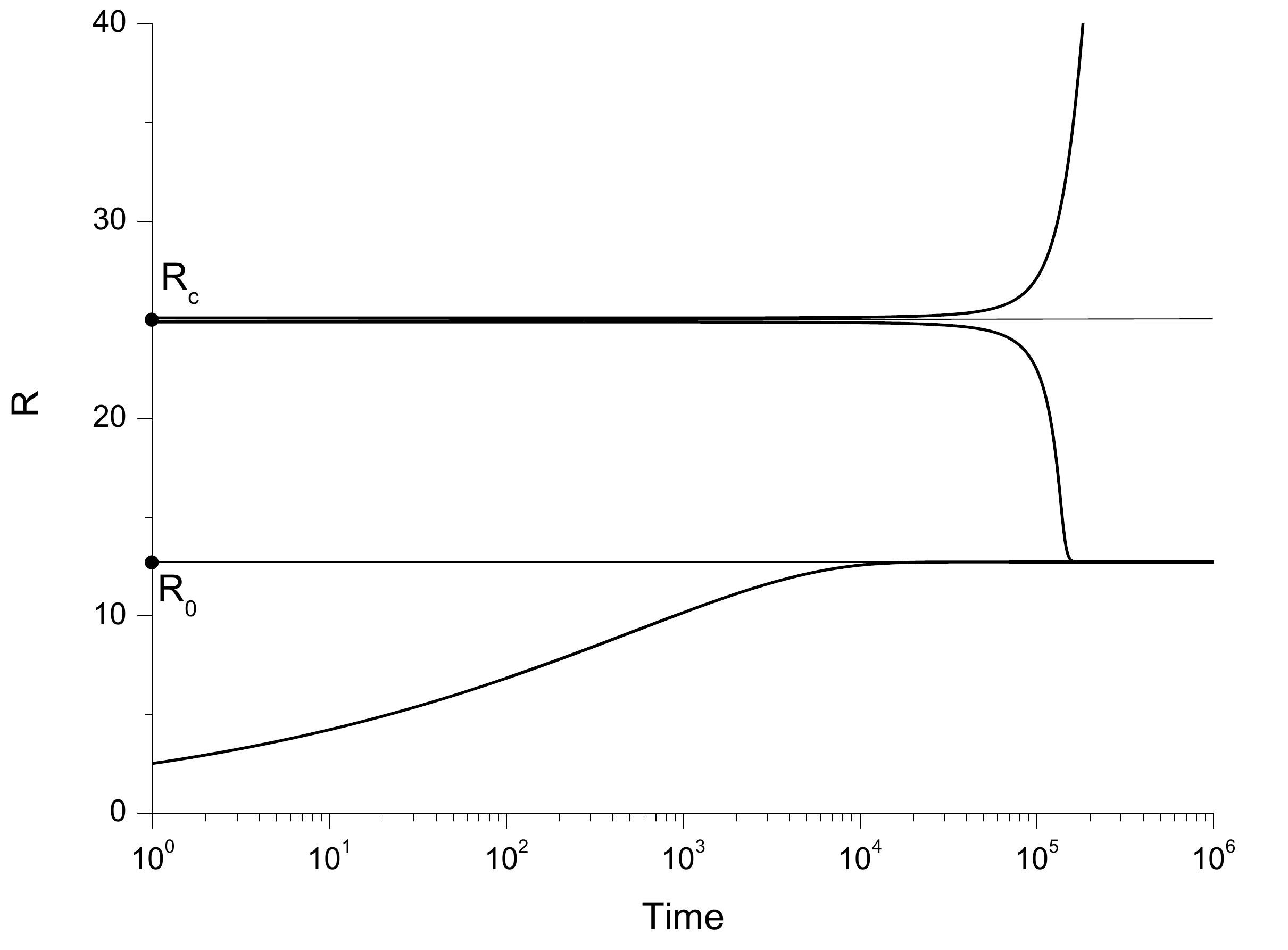}
 \caption{ Stationary values for $R_0$  at different values for $\varepsilon$  and $\mathcal{M}^{(1)}$ (plot (a)) and
 typical dependencies $R(t)$ at different initial conditions and at
$\alpha=0.6$, $\varepsilon=3$, $\mathcal{M}^{(1)}=0.5$ (plot (b)).
 Other parameters are: $D_0=3.0$, $\tau_J=0.3$, $\Delta=10^{-2}$.
  \label{rho_0t}}
\end{figure}

At first let us consider the simplest case of $\Delta=const$ where the critical
radius $R_c$ remains constant. In the stationary case we put $\dot R=0$ in
Eq.(\ref{rho}). Such assumption relates to autonomous regime of islands
formations. As far as $\Delta\sim 10^{-2}$ the critical radius is large but
finite. One of the stationary solution having physical meaning is $R_{0}=L$,
where $L\to \infty$ is the thermodynamic limit of the system size. There is
another solution $R_0$ responsible for formation of stationary patterns
characterized by $\Re\varkappa(\alpha,\varepsilon,\mathcal{M}^{(1)})\ne 0$. It
can be computed from the condition $R_{0}^5=G(x_m)(1+q^2R_{0}^2)^2$. The
corresponding dependencies $R_0(\alpha)$ at $R_0<R_c$ at different
$\varepsilon$ and $\mathcal{M}^{(1)}$ are shown in Fig.\ref{rho_0t}a. Its
values are bounded by stability criteria
$\Im\varkappa(\alpha,\varepsilon,\mathcal{M}^{(1)})\le 0$. It follows that
$R_0$ relate to large stable values for $x_m$\footnote{In the simplest case of
$x_m=0$ we arrive at classical results of the coalescence theory \cite{LLX}.}.
Using the potential $W(R)$ one can find that $W(R)$ has two asymptotics:
$W(R\to 0)\to \infty$, $W(R\to \infty )\to -\infty$. Therefore, the quantity
$R_{c}$ corresponds to a maximum of $W$, whereas $R_0$ defines its minimum
position. Hence, at $R>R_c$ all previously organized inclusions at early stages
start to grow; contrary, at $R<R_c$ the corresponding patterns relax to
spherical inclusions having fixed radius $R_0$. In order to illustrate the
corresponding time dependencies of the inclusion radius we solve numerically
Eq.(\ref{rho}) (see Fig.\ref{rho_0t}b).

As far as $R_0$ has unique dependence versus the system parameters, one can
solve Eq.(\ref{rho}) analytically rewriting it for the quantity $\rho\equiv
R/R_c$ in the form $\rho^3\dot \rho=(\rho-\rho_0)(\rho-1)$, where $\rho_0\equiv
R_0/R_c$, time derivative are taken with respect to $t'\equiv t/R_c^3$ (next we
drop prime for the time variable). Dividing variables, from the formal solution
of Eq.(\ref{rho}) following asymptotics can be found: in the vicinity of
$\rho=\rho_c\equiv 1$ there are exponentially increasing deviations from the
critical radius; in the vicinity of $\rho_0$ the exponential decaying $\rho(t)$
is observed.

Let us analyze the system behavior considering time dependence of the
supersaturation $\Delta(t)>0$. Hence there is a time dependent variable
$R_c(t)$. Next it will be convenient to move to new variables: $z(t)\equiv
R_c(t)/R_c(0)$, $u(t)\equiv R(t)/R_c(t)$. As a new time variable we introduce
$\tau=4\ln z$. It yields the derivative ${\rm d}u^4/{\rm
d}\tau=-u^4+\gamma(\tau)R^3{\rm d}R/{\rm d}t$, where
$\gamma(\tau)=(R_c^4(0)z^3{\rm d}z/{\rm d}t)^{-1}$. Using Eq.(\ref{rho}) and
taking into account that there is unique value $R_0$, one has $R^3{\rm d}R/{\rm
d}t=(R-R_0)(R/R_c-1)$. Therefore, an equation for the variable $u$ takes the
form
\begin{equation}\label{equ}
\frac{{\rm d}u^4}{{\rm d}\tau}=-u^4-R_0\gamma(u-1)+\gamma R_cu(u-1).
\end{equation}
Let us estimate time asymptotics for both $R(t)$ and $R_c(t)$. To this end we
consider the limit $\tau\gg 1$. It allows one to put ${\rm d}u^4/{\rm d}\tau=0$
or $u^4=-R_0\gamma(u-1)+\gamma R_cu(u-1)$. As far as the term in the left hand
side has no time dependent coefficient and $u(\tau\to \infty)=const$ it means
that all coefficients in the right hand side should be constants at $\tau\to
\infty$. When we assume $\gamma(\tau\to\infty)=const$ the coefficient $\gamma
R_c$ should take constant value too. Unfortunately, from the definition of
$\gamma(\tau)$ we arrive that $R_c(t)$ is an increasing function in time
leading to non-physical situation where all coefficients from the right hand
side of Eq.(\ref{equ}) are constants, not time-dependent functions. Therefore,
this assumption fails.

Let us consider another case, setting $\gamma R_c=const$ at $\tau\gg 1$.
Therefore, from the definition of $\gamma(\tau)$ one can find that
$R_c(t)\propto t^{1/3}$ and $\gamma(\tau)\to 0$. Therefore, the term
$R_0\gamma(u-1)$ can be neglected at $\tau\gg 1$. Using the definition of the
variable $u$ one can set that the radius of inclusion increases in time
according to the Lifshitz-Slyozov law $R(t)\propto t^{1/3}$ \cite{LS61}.

\section{Simulations}

In order to verify our analytical results we performed numerical simulations by
solving the system (\ref{eqPX}) in $d=2$-dimensional quadratic lattice of the
linear size $N=256$ with periodic boundary conditions with the mesh size
$\ell=0.5$ and the time step $\Delta t=0.001$. The noise term was modeled as
the Ornshtein-Uhlenbeck process \cite{Garcia}. As initial conditions we have
used $\langle x(\mathbf{r},0)\rangle=0.5$, $\langle(\delta x)^2\rangle=0.1$.
All measured quantities were averaged over 15 independent runs.

\begin{figure}
\centering
 a)\includegraphics[width=60mm]{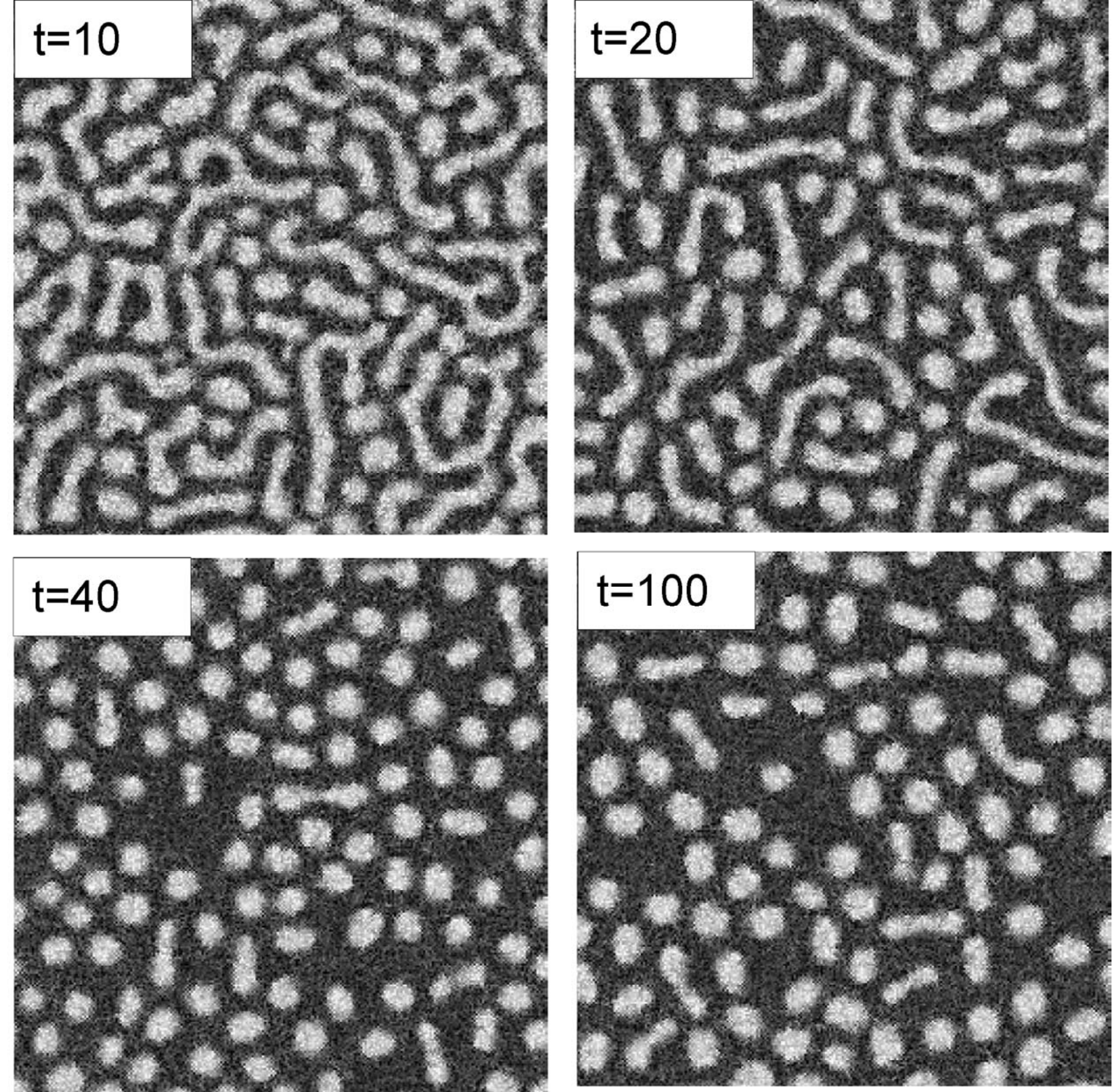}
 b)\includegraphics[width=80mm]{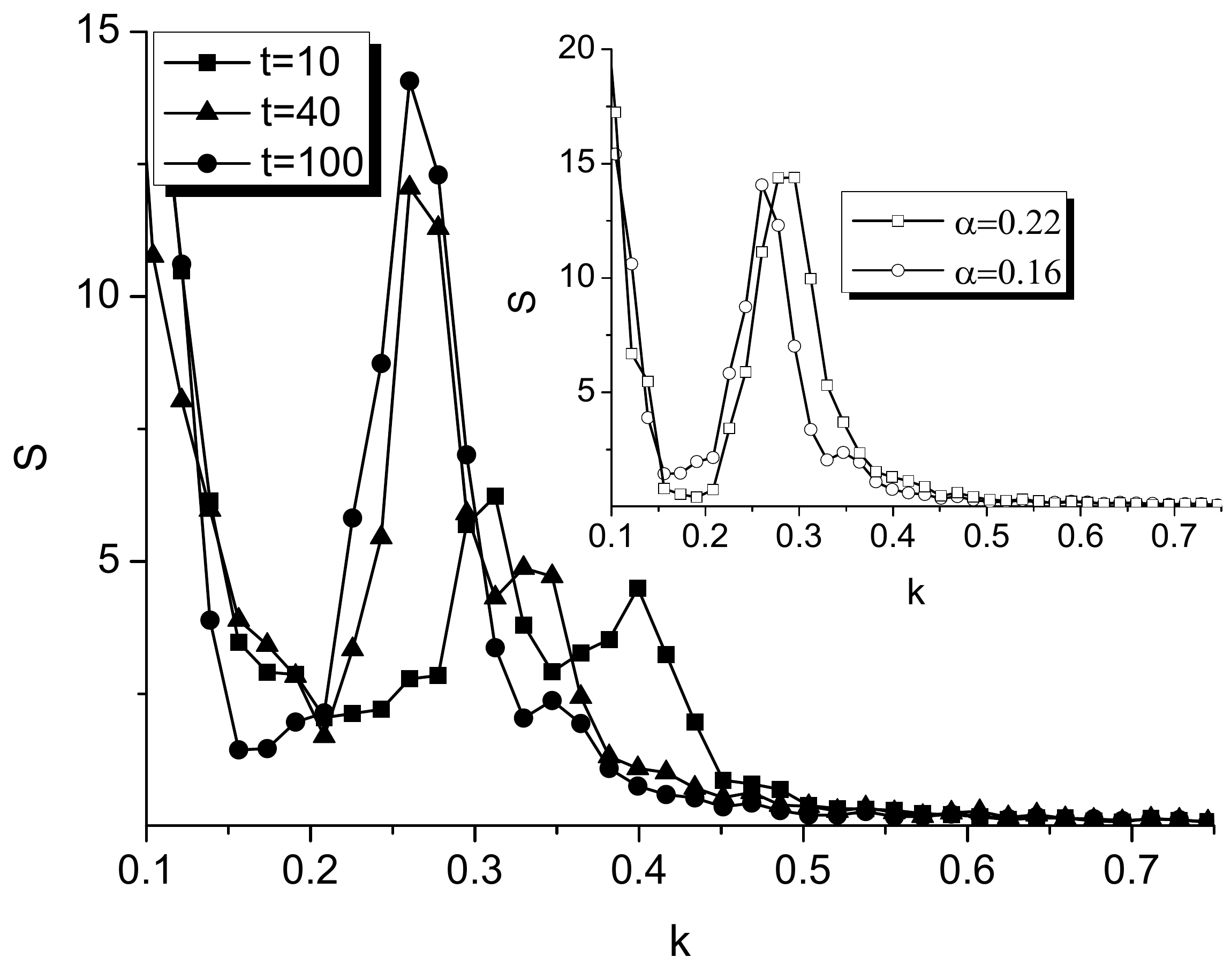}
\caption{Snapshots of the system evolution (a) and evolution of the spherically
averaged structure function (b) at $\alpha=0.16$. Insertion in plot (b) shows
structure function at $t=100$ and different values for adsorption rate. Other
parameters are: $D_0=3.0$, $\tau_J=0.3$, $\mathcal{M}^{(1)}=0.1$.
  \label{evol_pat}}
\end{figure}

Typical patterns organized during the system evolution are shown in
Fig.\ref{evol_pat}a at $\alpha=0.16$. It is seen that at small times patterns
of different sizes are formed. After the system selects patterns with a fixed
size. At large time intervals spherical patterns are organized having
preferably hexagonal symmetry as the Swift-Hohenberg operator predicts due to
elastic interactions \cite{EKHG02,EG04,Grant2006}. In Fig.\ref{evol_pat}b we
plot the corresponding spherically averaged structure function at different
times at relevant values for the wave-numbers to show the pattern selecting
processes. It is seen that except the main peak related to period of main
patterns there are well defined peaks at small wave-numbers (at elevated $k$
the corresponding peaks have very small intensity, not shown here). During the
system evolution its intensity decreases and only main peak is realized, its
position is shifted toward $\Re\varkappa$ defining the period of stationary
patterns. In insertion we have shown structure functions for two cases:
$\alpha=0.16$ and $\alpha=0.22$. At $\alpha=0.22$ the period of patterns is
smaller as the above stability analysis shows.

\begin{figure}
\centering
 a)\includegraphics[width=80mm]{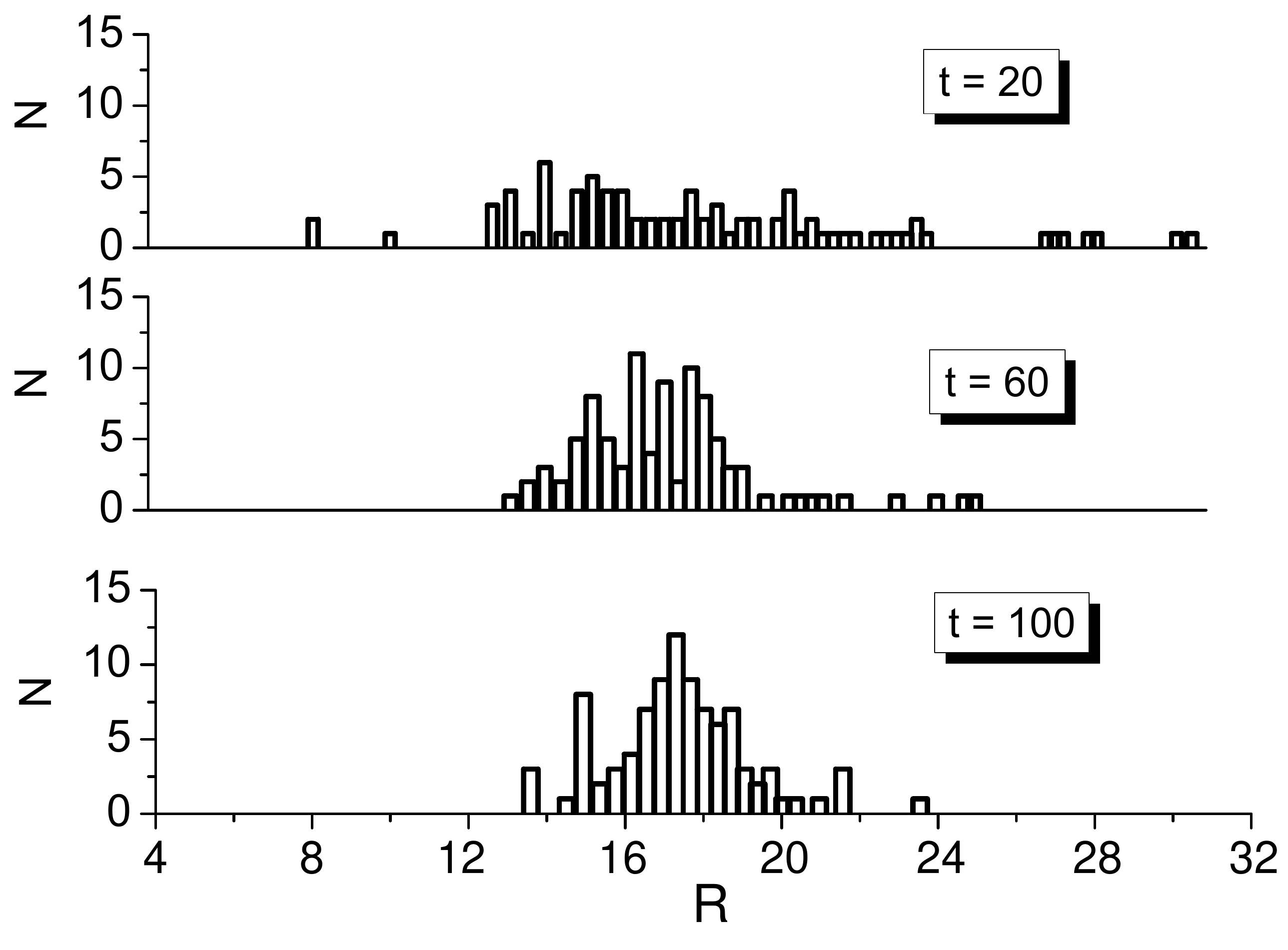} b)\includegraphics[width=80mm]{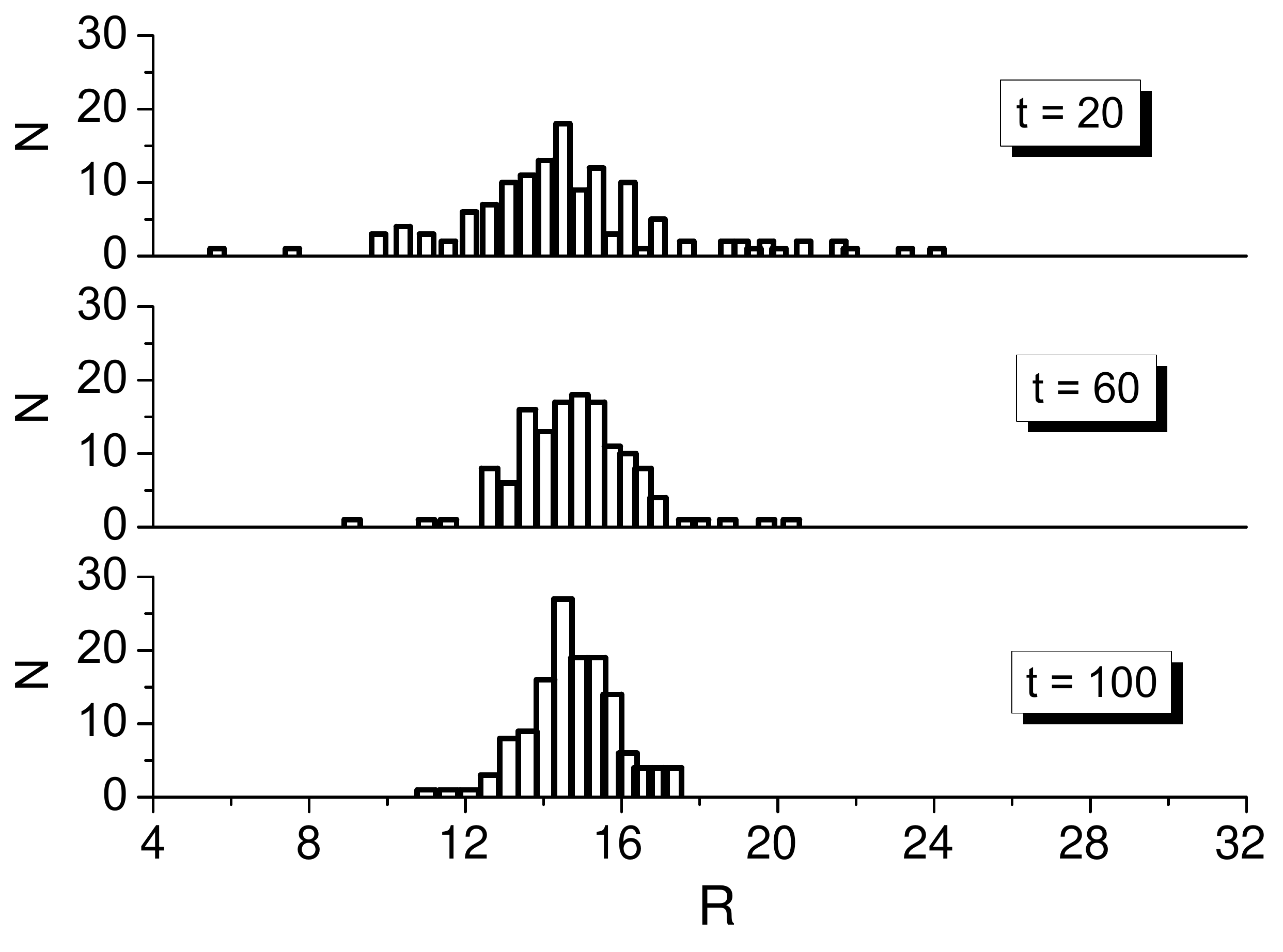}
\caption{Number of islands $N$ versus the island size $R$ at different times.
Plots (a) and (b) relate to $\alpha=0.16$ and $\alpha=0.3$
 Other parameters are: $D_0=3.0$, $\tau_J=0.3$, $\mathcal{M}^{(1)}=0.1$.
  \label{N_evol}}
\end{figure}

Considering quasi-stationary limit (time intervals where $\langle x\rangle$,
$\langle (x-\langle x\rangle)^2\rangle$ and $\langle (x-\langle
x\rangle)^3\rangle$ are time-independent quantities, $t\sim 10^2$) we can
compute radius of inclusions by measuring surfaces of islands (number of points
belonging to islands having spherical form). Evolution of the number of
inclusion having fixed radius $R$ at different $\alpha$ are shown in
Fig.\ref{N_evol}. It is seen that during the system evolution a distribution of
islands becomes more compact and it has one the most probable value for $R$.

\begin{figure}
\centering
 \includegraphics[width=80mm]{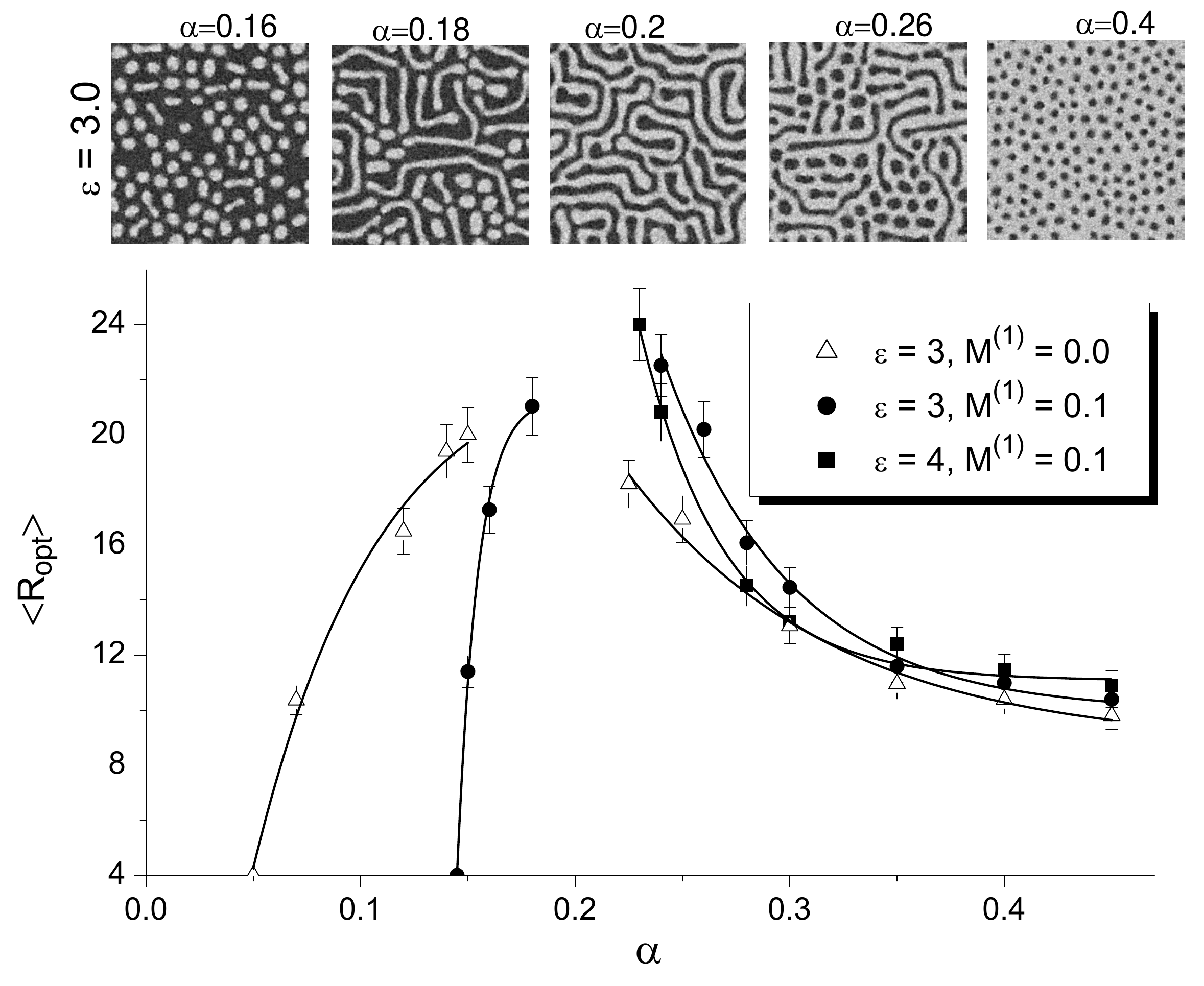}
\caption{Averaged linear size of islands versus adsorption rate $\alpha$ at
different $\varepsilon$ and $\mathcal{M}^{(1)}$. Stationary patterns observed
at $\varepsilon=0.3$ at different $\alpha$ are shown on the top. Other
parameters are: $D_0=3.0$, $\tau_J=0.3$.
  \label{rho_exp}}
\end{figure}

After averaging this radius over independent runs we obtain the dependencies of
the optimal characteristic length $\langle R_{opt}\rangle(\alpha)$ shown in
Fig.\ref{rho_exp}, where $\langle \cdots\rangle$ means average over
experiments. On the top we present snapshots of patterns realized at different
values for $\alpha$ at fixed $\varepsilon$. It is well seen that graphs
$\langle R_{opt}\rangle (\alpha)$ repeat analytical dependencies $R_0(\alpha)$
in Fig.\ref{rho_0t}a at small and large $\alpha$ where spherical islands
forming hexagonal crystal structure are possible. At intermediate values for
$\alpha$ strip patterns are realized and no spherical islands are possible.
Therefore, in Fig.\ref{rho_exp} there are gaps related to domains for $\alpha$
where no spherical islands are realized \footnote{In analytical study we have
assumed that all inclusions are of spherical form}. Fitting procedure allows us
to find that at small $\alpha$ where inclusions of the dense phase are possible
the exponential law is realized, $A-\langle R_{opt}\rangle=B^\alpha$, where $A$
and $B$ are fitting parameters. At large $\alpha$ when islands of vacancies are
realized one has exponential decaying, $\langle
R_{opt}\rangle-A=\exp(-\alpha/B)$. Therefore, an introduction of the
Swift-Hohenberg operator into the model for the system dynamics allows one to
describe formation of lattice with different kind of inclusions: at small
$\alpha$ one has molecular/atom inclusions forming hexagonal lattice, at large
$\alpha$ one has a situation when deposition of atoms on metal surface can be
described with formation of lattice of vacancy islands.

To estimate the spatial size of patterns one can consider typical limits of
$r_0\sim 1 nm$ and $L_{dif}\sim 1\mu m$. For the system of deposition of Al on
TiN (100) the estimation in Ref.\cite{CTT06} shows that at room temperature
with the lattice constant $a_{Al}=4.05\times 10^{-10}m$ and pair interaction
energy $\varepsilon=-0.22eV$ with lattice coordination number $Z=4$ the size of
patterns with $\mu=Za^2_{Al}\varepsilon/T$ is around $30\times 10^{-9}$m.
Therefore, patterns such as inclusions of dense/diluted phase are of
nano-scales and are \emph{nano-patterns} \cite{CTT07}. Patterns of diluted
phase usually are realized as lattice of vacancy islands (for example in
absorbed monoatomic layer of Ag deposited on Ru(0001) Ref.\cite{Nature99}).

\section{Conclusions}
We have studied pattern formation processes in stochastic reaction-diffusion
systems with the relaxation flux obeying Maxwell-Cattaneo equation. The novelty
of our approach is the usage of the phase field crystals formalism with special
geometry of realized patterns and introduction of multiplicative noise obeying
fluctuation-dissipation relation.

Pattern formation is discussed on the model system describing
adsorption-desorption processes. Considering pattern formation at early stages
of the system evolution we have analyzed temporal stability of patterns and
have shown that pattern selection processes are realized as a result of the
diffusion flux relaxation to its stationary value.

To study stationary patterns we have obtained the effective Fokker-Planck
equation and its stationary solution. It was shown that multiplicative noise
can induce non-equilibrium transitions at low energies for the adsorbate
interactions. Using stability analysis we have computed diagram illustrating
domains of the system parameters where stationary patterns can be realized. It
was shown that introduced multiplicative noise can induce re-entrant
pattern-forming transitions where stable/unstable stationary patterns exist
inside fixed intervals for lateral interaction energy of the adsorbate.
Stationary patterns organized during the system evolution form the crystalline
structure of hexagonal symmetry. It is explained by introduction of the
Swift-Hohenberg operator into the expansion for the interaction potential for
the adsorbate.

Considering quasi-stationary limit we have discussed evolution of radius of the
dense phase inclusions into diluted one or \emph{vice versa}. Using this
alternative approach related to coalescence theory it was shown that all
patterns are described by the characteristic radius $R_0$ that less then the
critical one $R_c$. In the vicinity of $R_c$ spatial perturbations grow
exponentially, whereas in the vicinity of $R_0$ such perturbations
exponentially decay. We have found that both critical and characteristic radii
have time asymptotics related to the Lifshits-Slyozov law.

\end{document}